\documentclass{ametsocV6.1}
\nolinenumbers
\let\internallinenumbers\relax
\usepackage[percent]{overpic}
\usepackage{siunitx}
\makeatletter
\def\p@subsection{\thesection}
\newcommand{\applabel}[2]{{\def\@currentlabel{#1}%
  \@ifundefined{phantomsection}{}{\phantomsection}\label{#2}}}
\makeatother

\title{RainFlow: Flow Matching for Infrared-Only Quantitative Precipitation Estimation}

\authors{
Matthieu Meignin,\aff{a,b}\correspondingauthor{Matthieu Meignin, matthieu.meignin@latmos.ipsl.fr}  
Cécile Mallet,\aff{a,b} and 
Nicolas Viltard\aff{a}
\thanks{This Work has been submitted to \textit{Artificial Intelligence for the Earth Systems}. Copyright in this Work may be transferred without further notice.}
}

\affiliation{
\aff{a}{LATMOS/IPSL, UVSQ Université Paris-Saclay, Sorbonne Université, CNRS, Guyancourt, France} \\
\aff{b}{ARCHES Team, Inria, Paris, France}
}

\abstract{ Geostationary infrared imagers observe precipitating systems every 10 to 15 minutes over large areas, but they sense cloud tops rather than the rain beneath. Their brightness temperatures constrain surface rain only weakly, so retrievals trained by regression return a smooth estimate that misses heavy rain and carries no measure of uncertainty. We treat infrared-only retrieval as conditional generation and introduce RainFlow, a flow-matching model that generates ensembles of rain fields from MSG/SEVIRI infrared observations, in a direct and a residual variant. Evaluated against radar over metropolitan France for a full year, the generated members are realistic rain fields, with the observed dry areas, heavy rain and spatial structure, and the ensemble mean is as accurate as a deterministic baseline. Used as probabilities, the ensembles locate heavy rain at scales where the baseline has no skill, and their spread captures most of the retrieval error, with only slight underdispersion. Trained with a time-sampling density that favours the noise end of the generative trajectory, the two variants reach similar skill. The direct variant depends strongly on this choice of density, whereas the residual variant is nearly insensitive to it. Evaluating the generated rain as a function of brightness temperature shows that the ensembles reproduce how rain depends on it over most of the range. It also isolates the remaining gap, an underestimate of both the amount and the variability of rain under the coldest cloud tops, which scores computed over all pixels do not reveal.
}

\begin{document}
\maketitle

\section{Introduction}
\label{sec:intro}

Quantitative precipitation estimation (QPE) at high spatial and temporal resolution is needed for hydrological modelling, flood warning and climate research, yet no single observing system can provide a reliable global product. Rain gauges measure rain directly but are sparse, unevenly distributed and confined to land. Weather radars resolve precipitating systems at kilometre scale, but cover only a small fraction of the land surfaces and almost none of the oceans. Passive microwave radiometers on low Earth orbit satellites respond to hydrometeors and provide the most physically direct measurement available from space, yet their revisit interval is long relative to the intermittence of rain. Geostationary infrared imagers offer the opposite trade-off: they observe their respective disc every ten to fifteen minutes, but sense mostly the temperature of cloud tops, which relates to surface rain only indirectly. However, this temporal continuity of measurements combined with the large field of view and the good spatial resolution offered by geostationary images makes these data a critical asset for QPE. Retrieving precipitation from geostationary infrared observations alone is the problem this study addresses.

The first infrared retrievals were built on the statistical association between cold cloud and rain. \citet{arkin1979} found that the fraction of an area covered by high, cold cloud explains most of the variance of rainfall accumulated over six hours, which led to estimating rainfall from the extent of cloud colder than a threshold \citep{arkin1987relationship}; PERSIANN-CCS \citep{hong2004ccs} later related the features of individual cloud patches to rain rate at the pixel scale. Deep convolutional networks now learn this relation directly from brightness temperature fields, trained under per-pixel losses against ground or spaceborne radar \citep{berthomier2023espresso, wang2024, meignin2026generative}. A model trained this way returns a summary statistic of the distribution of rain given the observations, its mean under a squared-error loss or its median under an absolute-error loss, rather than a sample from it. How much is lost depends on how tightly the observations constrain the rain field. From infrared observations alone, the constraint is weak: the retrieval is both an inversion, from cloud-top temperature to the hydrometeors that reach the ground, and a downscaling, from the roughly $6$\,km resolution of the imager over France to the $1$\,km of the radar reference, and infrared retrievals carry a large irreducible uncertainty \citep{amell2025roa}. We refer to this regime as weak conditioning throughout the paper. Under weak conditioning, the distribution of plausible rain fields is broad, and its mean is a smooth field that underestimates heavy rain \citep{berthomier2023espresso, meignin2026generative}.

Two lines of work move beyond this single estimate. Probabilistic retrievals estimate the distribution of rain rate at each pixel, for instance through quantile regression, and so report a case-specific uncertainty, from passive microwave \citep{viltard2023} to infrared observations \citep{amell2025roa}; because these distributions are estimated pixel by pixel, they do not yield spatially coherent rain fields. Conditional generative models sample the whole fields instead. Diffusion models have been used to generate ensembles of precipitation maps from combined microwave and infrared observations \citep{guilloteau2025difers} and to downscale atmospheric fields to kilometre scale \citep{mardani2025residual}. Flow matching has been used to super-resolve weather forecasts \citep{delefosse2026superresolving} and to produce probabilistic weather forecasts \citep{couairon2026archesweathergen}; \citet{rampal2026intercomparison} compare several generative families, including both, for precipitation downscaling. For infrared-only retrieval over France, \citet{meignin2026generative} compared deterministic and generative models on a common dataset. Deterministic models gave the more accurate pixel-wise estimates but underestimated heavy rain and smoothed the spatial variability of rain, whereas generative models, a conditional GAN and a diffusion model, better reproduced that variability and the heavy tail of the distribution. That study also identified the cost of diffusion sampling, which requires many iterative denoising steps, as a main obstacle to operational use, and pointed to flow matching, which generates samples in fewer iterations, as a remedy. It did not, however, evaluate whether the spread of the generated ensembles matches the retrieval error. We build on that study by adopting flow matching, and use it to examine two design choices that remain largely unexamined. First, several generative models for downscaling and forecasting produce a residual around a deterministic estimate rather than the field itself \citep{mardani2025residual, couairon2026archesweathergen}, usually without a controlled comparison with direct generation, whereas infrared retrieval has so far used direct generation \citep{meignin2026generative}. Second, the density from which training times are drawn is often taken from image generation \citep{esser2024scaling}, although it sets where along the generative trajectory the network concentrates its training.

A generative retrieval returns an ensemble of members, and evaluating it raises questions that a single estimate does not. Existing studies use several diagnostics. Statistics of the field, such as the distribution of rain intensity and the power spectrum, check that members are realistic \citep{guilloteau2025difers}. The continuous ranked probability score measures the skill of the ensemble \citep{couairon2026archesweathergen, delefosse2026superresolving, rampal2026intercomparison}, neighbourhood scores assess where the rain is located \citep{roberts2008scale}, and the spread--skill ratio evaluates the dispersion of the ensemble \citep{fortin2014should}. All of these diagnostics are organised around the observed rain. None of these evaluations of generative retrievals asks whether, at a given brightness temperature, the retrieved rain has the amount and variability of the observed rain. Under weak conditioning, this is not guaranteed, especially at brightness temperatures with few training examples.

This study addresses these questions for infrared-only retrieval over metropolitan France, building on the dataset and deterministic baseline of \citet{meignin2026generative}: seven MSG/SEVIRI infrared and water-vapour channels, with no microwave, gauge or numerical-weather-prediction input, are used to retrieve instantaneous rain rate at $1$\,km, evaluated against the M\'et\'eo-France radar composite over a full withheld year. The study makes three main contributions. We introduce RainFlow, to our knowledge the first flow-matching models for infrared-only precipitation retrieval \citep{lipman2023flow, tong2024improving}, in a direct and a residual variant, and show that the time-sampling density matters more than the choice between them: with a density favouring the noise end, both reach similar skill, but the direct variant is far more sensitive to this choice. We assemble an evaluation protocol for generative retrievals that makes explicit what is scored, from the ensemble mean and individual members to the pooled members and the full ensemble, and that adds an evaluation of the retrieval as a function of the infrared conditioning. Applying it, we find that the generated members are realistic rain fields at no cost in mean accuracy, that the ensembles provide skilful exceedance probabilities for heavy rain at scales where the deterministic baseline has none, with a mildly underdispersed spread, and that the remaining limitations are concentrated in deep convection under the coldest cloud tops, a regime that is rare in the training data.

Section~\ref{sec:data} describes the data, Section~\ref{sec:method} the two models and the objective they share, and Section~\ref{sec:results} the diagnostics and their results. Section~\ref{sec:discussion} discusses what these results reveal about generative retrieval under weak conditioning, and Section~\ref{sec:conclusion} concludes.

\section{Data}
\label{sec:data}

We use the dataset of \citet{meignin2026generative} unchanged; all details of its construction are given there. Each sample pairs one MSG/SEVIRI acquisition with the co-located M\'et\'eo-France radar mosaic over metropolitan France, on tiles of $256\times256$\,\si{\kilo\metre\squared}, that is, $256\times256$ pixels on the $1$\,km reference grid.

\paragraph{Ground truth.} The reference is the surface rain rate from the quality-controlled M\'et\'eo-France operational radar composite, at $1$\,km and $5$\,min, giving a rain rate field $RR \in \mathbb{R}^{1\times256\times256}$ in \si{\milli\metre\per\hour}. The composite is quantised in steps of $\SI{0.12}{\milli\metre\per\hour}$, with no value between zero and that step, so a pixel is counted as dry when $RR < \SI{0.12}{\milli\metre\per\hour}$, for the observations and every model alike.

\paragraph{Infrared satellite observations.} The conditioning $c$ consists of seven MSG/SEVIRI infrared and water-vapour channels ($6.2$, $7.3$, $8.7$, $9.7$, $10.8$, $12.0$ and $13.4$\,\si{\micro\metre}), referred to by type and wavelength (e.g.\ $\mathrm{IR}_{10.8}$, $\mathrm{WV}_{6.2}$). Together they constrain cloud-top temperature and height, cloud phase and upper-tropospheric water vapour. SEVIRI has a nominal resolution of $3$\,km at nadir, which the off-nadir viewing geometry degrades to about $6$\,km over metropolitan France \citep{eumetsat2017}, and a repeat cycle of $15$\,min. Each scene is upsampled to the $1$\,km grid of the reference and matched with the nearest radar frame, giving $c \in \mathbb{R}^{7\times256\times256}$. 

\paragraph{Domain and splits.} Tiles are drawn from 2008 to 2023, with the full year 2019 held out for test and separate days for validation. Because precipitation, and heavy precipitation in particular, is rare and therefore hard to learn, the dataset is oversampled toward precipitating scenes and heavy precipitation, as is common in deep learning for imbalanced data, identically across splits. Aggregate frequencies such as the dry fraction (\S\ref{sec:res-dist}), the marginal $P(RR)$ (Fig.~\ref{fig:marginal}) and the false-alarm ratio (Table~\ref{tab:em-pixel}) therefore describe the evaluation set rather than the operational climatology, and are compared between models rather than taken as absolute values; validation under natural sampling is a separate exercise.

\section{Method}
\label{sec:method}

We treat quantitative precipitation estimation from infrared observations as conditional generation: rather than predicting a single field, we learn to sample from $p(RR \mid c)$, with $RR$ the rain rate field and $c$ the infrared conditioning. Under the weak conditioning described in \S\ref{sec:intro}, this distribution is broad, and its mean, which a model trained under a squared-error loss returns, is a smooth field resembling none of the plausible ones. The same framing is adopted by \citet{guilloteau2025difers}, who generate ensembles of precipitation maps by diffusion from combined microwave and infrared observations; here we work from infrared alone and retrieve instantaneous rain rate.

We use conditional flow matching (CFM) and compare two ways of setting up what the generative model produces, both evaluated against a deterministic baseline. The baseline is the multitask UNet of \citet{meignin2026generative}, trained to predict both the rain rate, with a mean-squared-error loss, and the occurrence of rain, with a binary cross-entropy loss; it provides a smooth best estimate of the rain field and is kept fixed here. Both generative models receive the infrared conditioning $c$. The first, RainFlow, generates the rain field directly. The second, RainFlow-Residual, also receives the baseline prediction and generates the variability around it. Section~\ref{sec:flow-matching} presents the flow-matching objective shared by both variants, \S\ref{sec:direct} and \S\ref{sec:two-stage} what each one generates, \S\ref{sec:time-sampling} the density from which training times are drawn, and \S\ref{sec:training-sampling} their training and the generation of ensembles.

\subsection{Conditional flow matching}
\label{sec:flow-matching}

We train with conditional flow matching \citep{lipman2023flow, tong2024improving} under the linear probability path and the independent coupling, in which the source sample is drawn independently of the target rather than paired with it by minibatch optimal transport \citep{tong2024improving}. Let $x_1$ be a target field, $x_0 \sim \mathcal{N}(0, I)$ an independently drawn source sample, and $t \in [0,1]$ a time, with the convention that $t=0$ is noise and $t=1$ is data. Each pair $(x_0, x_1)$ defines a conditional path $x_t$ and a conditional velocity $u_t$,

\begin{equation}
x_t = t\,x_1 + (1-t)\,x_0,
\qquad
u_t(x_t \mid x_0, x_1) = \frac{\mathrm{d}x_t}{\mathrm{d}t} = x_1 - x_0 ,
\label{eq:cond-path}
\end{equation}

a straight line traversed at constant velocity, as in rectified flow \citep{liu2022flowstraightfastlearning} and in the zero-noise case of the stochastic-interpolant framework \citep{albergo2025stochasticinterpolantsunifyingframework}. With $x_0 \sim \mathcal{N}(0, I)$ and $x_1 \sim p(x_1 \mid c)$, the distribution of $x_t$ defines the marginal probability path $p_t(\cdot \mid c)$, which goes by construction from the Gaussian prior $\mathcal{N}(0, I)$ at $t=0$ to the data distribution $p(x_1 \mid c)$ at $t=1$. Generating a sample amounts to transporting a Gaussian draw along this path. Here $x_1$ is not the rain rate $RR$ itself but its transform to model space, defined in \S\ref{sec:direct} for direct generation and in \S\ref{sec:two-stage} for the residual variant.

The network $v_\theta$, with parameters $\theta$, takes the current state, the time and the infrared conditioning $c$, and is trained by regression on the conditional velocity of the pair that produced that state:

\begin{equation}
\mathcal{L}_{\mathrm{CFM}}(\theta) = \mathbb{E}_{t \sim \mathcal{U}[0,1],\; x_0,\; (x_1, c)}
\Big\lVert v_\theta(t, x_t, c) - (x_1 - x_0) \Big\rVert_2^2 ,
\label{eq:cfm-loss}
\end{equation}

where $t$ is drawn uniformly as in the original formulations \citep{lipman2023flow, tong2024improving}; the modified density used in this work is given in \S\ref{sec:time-sampling}.

The conditional paths of different pairs intersect \citep{liu2022flowstraightfastlearning}: a given state, reached at time $t$, lies on the paths of many pairs, each with its own velocity $x_1 - x_0$. The network sees only the state, the time and the conditioning, not the pair that produced the state, so it cannot recover the velocity of each pair; under a squared-error loss, the best it can predict is their average. For every time $t$ and state $x$, the minimiser of \eqref{eq:cfm-loss} is therefore the expected velocity of the pairs whose path passes through $x$ at time $t$ \citep{liu2022flowstraightfastlearning},

\begin{equation}
\bar{u}_t(x \mid c) = \mathbb{E}\big[\, x_1 - x_0 \;\big|\; x_t = x,\, c \,\big] ,
\label{eq:marginal-velocity}
\end{equation}

the marginal velocity field, which generates the path $p_t(\cdot \mid c)$ \citep{lipman2023flow}. The network $v_\theta$ is an estimator of this field.

A sample is generated by integrating the learned field from a Gaussian draw,

\begin{equation}
\frac{\mathrm{d}x_t}{\mathrm{d}t} = v_\theta(t, x_t, c), \quad x_0 \sim \mathcal{N}(0, I),
\qquad \text{so that} \qquad
x_1 = x_0 + \int_0^1 v_\theta(t, x_t, c)\,\mathrm{d}t ,
\label{eq:sampling-ode}
\end{equation}

where $x_t$ now denotes the trajectory of the learned flow. Up to estimation and integration error, $x_1$ is then a sample from $p(x_1 \mid c)$, which the inverse transforms of \S\ref{sec:direct} and \S\ref{sec:two-stage} map back to a rain rate field $RR$. The integral is evaluated numerically; the solver and the ensemble size are given in \S\ref{sec:training-sampling}.

The two variants below share this objective, network architecture and solver. They differ only in what $x_1$ represents, the rain field itself (\S\ref{sec:direct}) or the variability around the baseline prediction (\S\ref{sec:two-stage}), and in what the network is conditioned on.

\subsection{RainFlow: direct generation}
\label{sec:direct}

RainFlow generates the rain field itself. Rain rate is heavy-tailed and dominated by dry pixels, so, as is common for precipitation, it is mapped to model space through the log rain rate $\widetilde{RR} = \log(1 + RR)$, followed by a standardisation,

\begin{equation}
x_1 = \frac{\widetilde{RR} - m}{\sigma},
\label{eq:model-space}
\end{equation}

where $m$ and $\sigma$ are the mean and standard deviation of $\widetilde{RR}$ over the training split. The target $x_1$ is thus a normalised log rain rate: the transform compresses the heavy tail and brings the target to the unit scale of the Gaussian source, while remaining exactly invertible. The infrared conditioning $c$, the seven infrared and water-vapour channels of \S\ref{sec:data}, is standardised per channel over the same split.

The network receives $x_t$ and $c$ as an eight-channel input and returns a single velocity channel. A sample $x_1$, obtained by integrating \eqref{eq:sampling-ode}, is mapped back to rain rate by inverting \eqref{eq:model-space},

\begin{equation}
RR = \exp\!\big(\sigma x_1 + m\big) - 1 .
\label{eq:invert-direct}
\end{equation}

\subsection{RainFlow-Residual: residual generation}
\label{sec:two-stage}

Where RainFlow generates the rain field itself, RainFlow-Residual generates the variability around an existing estimate: the prediction $RR_{\mathrm{det}}$ of the baseline of \citet{meignin2026generative}, which is smooth and under-predicts extremes because it estimates a conditional mean. In log rain rate, the field is split into the part a regression can predict and a residual,

\begin{equation}
\widetilde{RR} \;=\; \underbrace{\mathbb{E}\big[\widetilde{RR} \mid c\big]}_{\textstyle :=\, \mu \;\text{(regression)}}
\;+\; \underbrace{\widetilde{RR} - \mathbb{E}\big[\widetilde{RR} \mid c\big]}_{\textstyle :=\, r \;\text{(residual)}} .
\label{eq:two-stage-split}
\end{equation}

The regression term $\mu$ is approximated by the log of the baseline prediction, $\widetilde{RR}_{\mathrm{det}} = \log(1 + RR_{\mathrm{det}}) \simeq \mu$. Because the baseline is trained on rain rate rather than on its logarithm, this is not strictly the conditional mean of the log field, but the mean of the residual remains close to zero in practice. The flow-matching model of \S\ref{sec:flow-matching} generates the residual $r$, scaled to unit variance,

\begin{equation}
x_1 \;=\; \frac{\widetilde{RR} - \widetilde{RR}_{\mathrm{det}}}{\sigma_r} \;\simeq\; \frac{r}{\sigma_r},
\label{eq:two-stage-target}
\end{equation}

with $\sigma_r$ the standard deviation of $\widetilde{RR} - \widetilde{RR}_{\mathrm{det}}$ over the training split, so that $x_1$ is a normalised residual. Splitting the target into a regressed mean and a generated residual follows \citet{mardani2025residual, guilloteau2025difers, rampal2026intercomparison, couairon2026archesweathergen}.

The network receives $x_t$, $c$ and $\widetilde{RR}_{\mathrm{det}}$ as a nine-channel input, so the residual it generates can depend on the baseline prediction. A sample $x_1$, obtained by integrating \eqref{eq:sampling-ode}, is mapped back to rain rate by inverting \eqref{eq:two-stage-target},

\begin{equation}
RR = \big(1 + RR_{\mathrm{det}}\big)\,\exp\!\big(\sigma_r x_1\big) - 1 .
\label{eq:invert-two-stage}
\end{equation}

\subsection{Time-sampling density}
\label{sec:time-sampling}

The two variants thus differ in what they generate but share the objective \eqref{eq:cfm-loss}, written there with $t$ drawn uniformly. This density is in fact a free choice: since the marginal velocity \eqref{eq:marginal-velocity} minimises the regression error at each time separately, it remains the minimiser for any density covering $[0,1]$. What the density changes is how a network of finite capacity, trained for a finite number of steps, divides its effort along the trajectory. \citet{esser2024scaling} obtain their best overall results with a logit-normal density, which concentrates in the interior of the interval and vanishes at both ends, and which has since been widely adopted, including for atmospheric fields \citep{delefosse2026superresolving,couairon2026archesweathergen}.

Under the independent coupling, the two ends of the trajectory make very different use of the conditioning. At $t = 0$ the state is the source draw, independent of the target, and at $t = 1$ it is the target itself, so that \eqref{eq:marginal-velocity} reduces to
\begin{equation}
\bar{u}_0(x \mid c) = \mathbb{E}[x_1 \mid c] - x,
\qquad
\bar{u}_1(x \mid c) = x .
\label{eq:endpoint-velocities}
\end{equation}
Near the data end, the velocity is set by the state alone and the conditioning is redundant. Near the noise end, it is set by the conditional mean of the target, which the network can only learn from the conditioning. Without a conditioning, this target would reduce to the mean of the data, a constant, and both ends would be trivial: \citet{lee2024improving} note that the useful part of training then lies in the middle of the interval, where the logit-normal density concentrates. Under weak conditioning, the noise end is no longer trivial: the conditional mean, the amplitude of the rain field given the infrared observations, must be learned from a weak signal, at the start of the generative trajectory, the end that the uniform density does not favour and the logit-normal density visits least.

This matters far more for one variant than for the other. RainFlow must recover $\mathbb{E}[x_1 \mid c]$, the normalised conditional mean of the log rain rate, from the infrared conditioning alone. For RainFlow-Residual, the baseline prediction, which the network receives as input, already supplies it: the residual is nearly centred in practice (\S\ref{sec:two-stage}), so that $\mathbb{E}[x_1 \mid c] \simeq 0$ and little remains to be learned at the noise end.

We therefore draw the training time from a power-law density that favours the noise end, $t = u^{k}$ with $u \sim \mathcal{U}[0,1]$, and set $k = 3$ by a grid search over $k \in \{1.5, 3, 5\}$. This places $46.4\%$ of draws below $t = 0.1$, against $10\%$ for uniform sampling and $1.4\%$ for the logit-normal density (Fig.~\ref{fig:tdensity}). The same density is used for both variants in all experiments of \S\ref{sec:results}; Appendix~\ref{app:tsampling} compares it with the uniform and logit-normal densities, and \S\ref{sec:res-ssr} summarises the result.

\subsection{Training and ensemble generation}
\label{sec:training-sampling}

With the objective and its time density fixed, what remains is the network that estimates the velocity field, its training, and the generation of the ensembles evaluated in \S\ref{sec:results}; these choices are common to both variants and follow standard practice for flow-matching models. The velocity network $v_\theta$ uses the UNet architecture of \citet{dhariwal2021adm} at $256 \times 256$:  base width $64$, channel multipliers $(1,1,2,3,4)$ across five resolutions, two residual blocks per resolution, self-attention with four heads of $64$ channels at the $16 \times 16$ resolution, and no dropout. The state $x_t$ and the conditioning are concatenated at the input layer (\S\ref{sec:direct}, \S\ref{sec:two-stage}), and the output is a single velocity channel. Time enters through a sinusoidal embedding projected by a two-layer MLP, which modulates every residual block through a per-channel scale and shift of its normalised features: a feature-wise linear modulation (FiLM; \citealp{perez2018film}), implemented by adaptive group normalisation \citep{dhariwal2021adm}.

This network is trained on the objective \eqref{eq:cfm-loss}, with $t$ drawn from the density of \S\ref{sec:time-sampling}, using AdamW at learning rate $10^{-4}$ and no weight decay, batch size 16, for 400{,}000 steps, with 5{,}000 steps of linear warmup followed by cosine decay to $0.1\times$ peak and gradients clipped to global norm 1.0. An exponential moving average of the weights with decay 0.9999 is maintained from the midpoint of training, and all reported results use the averaged weights. We select the checkpoint minimising \eqref{eq:cfm-loss} on the full validation split, evaluated every 5{,}000 steps with the same random timesteps, so that successive values differ only through the weights. A sample-based score such as the CRPS (App.~\ref{app:evaluation-metrics}) would be a better criterion, but it requires an ODE solve per validation tile, which is too costly at this frequency.

Once trained, the network generates samples by integrating \eqref{eq:sampling-ode} with the adaptive Dormand--Prince solver \texttt{dopri5} \citep{DORMAND198019,chen2018neuralode} at tolerance $10^{-3}$, and an ensemble of fifty members is obtained from fifty independent source draws $x_0$ for the same conditioning.

\clearpage

\section{Results}
\label{sec:results}

For each of the $1349$ test scenes of 2019, RainFlow and RainFlow-Residual each generate an ensemble of $50$ members. An ensemble can be evaluated in several ways, each answering a different question, and we use five of them throughout this section (Table~\ref{tab:notation}). The ensemble mean (EM) averages the members into a single field; like the deterministic baseline, it estimates the conditional mean of the rain rate, and we use it to check that the ensembles are correctly centred. A random member (RM) is a single generated field, used to judge whether individual samples are realistic. The average member (AM) gives the scores of a typical member: each member is scored on its own and the scores are averaged over the $50$, so that they do not depend on a particular draw. Pooling the 50 members over all scenes (50M) gives one large sample of the joint distribution of brightness temperature and rain rate, from which we derive the moments of the rain rate in each brightness temperature bin (Fig.~\ref{fig:condtb}) and the distribution of brightness temperature in each rain rate bin (Fig.~\ref{fig:ptbr}). The full ensemble (E) is scored as a probabilistic prediction, for instance with the continuous ranked probability score. Since the ensemble mean behaves like the baseline on most diagnostics, it is shown only where it adds information.

\begin{table}[t]
  \centering
  \small
  \setlength{\tabcolsep}{3pt}
  \begin{tabular}{@{}l p{0.2\linewidth} p{0.45\linewidth} p{0.10\linewidth}@{}}
    \hline
    Notation & Object evaluated & Question answered & Sections \\
    \hline
    EM  & mean of the members                 & Is the ensemble correctly centred? & \S\ref{sec:res-pixel} \\
    RM  & one random member                   & Is a single generated field realistic? & \S\ref{sec:res-dist} \\
    AM  & each member scored, scores averaged & How does a typical member score? & \S\ref{sec:res-pixel}, \S\ref{sec:res-dist}, \S\ref{sec:res-realism} \\
    50M & members pooled                      & Is the generated distribution consistent with the conditioning? & \S\ref{sec:res-cond} \\
    E   & full ensemble                       & Is the ensemble skilful and calibrated? & \S\ref{sec:res-realism}, \S\ref{sec:res-ssr} \\
    \hline
  \end{tabular}
  \caption{\textbf{Five ways of evaluating an ensemble of $50$ members.} AM scores are reported as mean $\pm$ standard deviation over the members.}
  \label{tab:notation}
\end{table}

\subsection{Recovering the conditional mean}
\label{sec:res-pixel}

\begin{table}[htbp]
  \centering
  \small
  \begin{tabular}{lccccc}
    \hline
    Model & RMSE $\downarrow$ & MAE $\downarrow$ & CSI $\uparrow$ & POD $\uparrow$ & FAR $\downarrow$ \\
    \hline
    Baseline               & \textbf{1.484} & 0.408          & \textbf{0.559} & 0.884          & \textbf{0.397} \\
    \hline
    RainFlow (EM)          & 1.495          & \textbf{0.400} & 0.556          & 0.887          & 0.402 \\
    RainFlow-Residual (EM) & 1.487          & 0.408          & 0.551          & \textbf{0.891} & 0.409 \\
    \hline
  \end{tabular}
  \caption{\textbf{Pixel-wise skill of the ensemble mean on the 2019 test set.} Baseline against the 50-member ensemble mean (EM) of each generative model. RMSE and MAE in \si{\milli\metre\per\hour}; CSI, POD and FAR for the rain/no-rain event $RR \geq \SI{0.12}{\milli\metre\per\hour}$. Best per column in bold; arrows give the preferred direction.}
  \label{tab:em-pixel}
\end{table}

We first check that the ensembles are correctly centred. A member is a draw from $p(RR \mid c)$, the distribution of rain given the infrared conditioning $c$, rather than an estimate of its expectation, so it cannot improve on that estimator under a squared-error score, and its pixel-wise scores (AM) are correspondingly poorer (Table~\ref{tab:pixel}). The ensemble mean estimates $\mathbb{E}[RR \mid c]$, as does the baseline, which is trained by regression in the space in which RMSE is evaluated and is therefore near-optimal on that score by construction. The ensemble mean should thus approach it from above, and Table~\ref{tab:em-pixel} is best read as a consistency check.

The ensemble means meet this expectation. On RMSE, neither improves on the baseline, but both come within $1\%$ of it: $1.487$ for RainFlow-Residual (EM) and $1.495$ for RainFlow (EM), against $1.484$. On MAE, RainFlow (EM) does slightly better ($0.400$ against $0.408$), and RainFlow-Residual (EM) matches the baseline.

Relative to the baseline, the ensemble means detect slightly more of the observed rain (higher POD) at the cost of slightly more false alarms (higher FAR), which leaves CSI nearly unchanged. This is the expected behaviour of an ensemble mean: individual members put precipitation cells in slightly different locations, so averaging spreads low-amplitude precipitation over the union of plausible locations. Single members (AM) show the mirror image: committing to one location, they detect less of the observed rain (POD around $0.60$) but raise fewer false alarms (FAR around $0.35$; Table~\ref{tab:pixel}, App.~\ref{app:member-pixel}).

Across all five scores, the ensemble means barely differ from the baseline, and the two models differ from each other by at most $2\%$: the ensembles are correctly centred. This is a necessary condition for the calibration analysis that follows, and it means that the ensemble mean and the baseline can be used interchangeably as estimators of $\mathbb{E}[RR \mid c]$. We therefore take the baseline as the reference conditional-mean estimator in the rest of this paper. The value of generative models lies not in their average but in their members. The next question is therefore whether these members follow the distribution of the observed rain.

\subsection{Distributional fidelity}
\label{sec:res-dist}

Rain rate is difficult to reproduce: in our dataset, most pixels are dry, and above zero the distribution is heavy-tailed, spanning more than three orders of magnitude, from $0.12$ to about $\SI{300}{\milli\metre\per\hour}$. Under weak conditioning, a conditional-mean estimator cannot represent either feature, since averaging over plausible fields smooths away dry gaps and heavy rain alike. This is where generative models make the clearest difference: their members reproduce the dry fraction, the full rain rate distribution and the spatial structure of the observed fields.

\begin{table}[htbp]
  \centering
  \small
  \begin{tabular}{lc}
    \hline
    Model & Dry fraction (\%) \\
    \hline
    Ground Truth             & $71.40$ \\
    \hline
    Baseline                 & $58.08$ \\
    \hline
    RainFlow (AM)            & $73.71 \pm 0.19$ \\
    RainFlow-Residual (AM)   & $\mathbf{73.06 \pm 0.18}$ \\
    \hline
  \end{tabular}
  \caption{\textbf{Dry fraction of single members on the 2019 test set.} Fraction of dry pixels, defined as $RR < \SI{0.12}{\milli\metre\per\hour}$. For the generative models, the fraction is computed for each single member and reported as mean $\pm$ standard deviation over the $50$ members (AM). Closest to the observations in bold.}
  \label{tab:dry}
\end{table}

\paragraph{Dry fraction.} Rain is intermittent, and the fraction of dry pixels measures how well this intermittence is reproduced. Here, as throughout, a pixel is dry below $\SI{0.12}{\milli\metre\per\hour}$, the quantisation step of the radar. Averaged over the $50$ members (AM), a member is dry on $73.7\%$ of pixels for RainFlow and $73.1\%$ for RainFlow-Residual, against $71.4\%$ observed, and this fraction is nearly identical from one member to the next (Table~\ref{tab:dry}). The baseline is dry on only $58.1\%$ of pixels: it reports rain over $42\%$ of the domain where $29\%$ is observed. This follows from what a conditional mean is. At a pixel where only some of the plausible fields are wet, the conditional mean is a light rain weighted by the probability of rain, so it is wet wherever rain is plausible, and its rain area approaches the union of the plausible rain areas rather than the area of any one of them. A member instead samples, at each location, whether it rains, and so reproduces the rain area of a single field; its slight excess of dry pixels is consistent with the lower detection rate of single members (\S\ref{sec:res-pixel}). 

\begin{figure}[htbp]
  \centering
  \includegraphics[width=\linewidth]{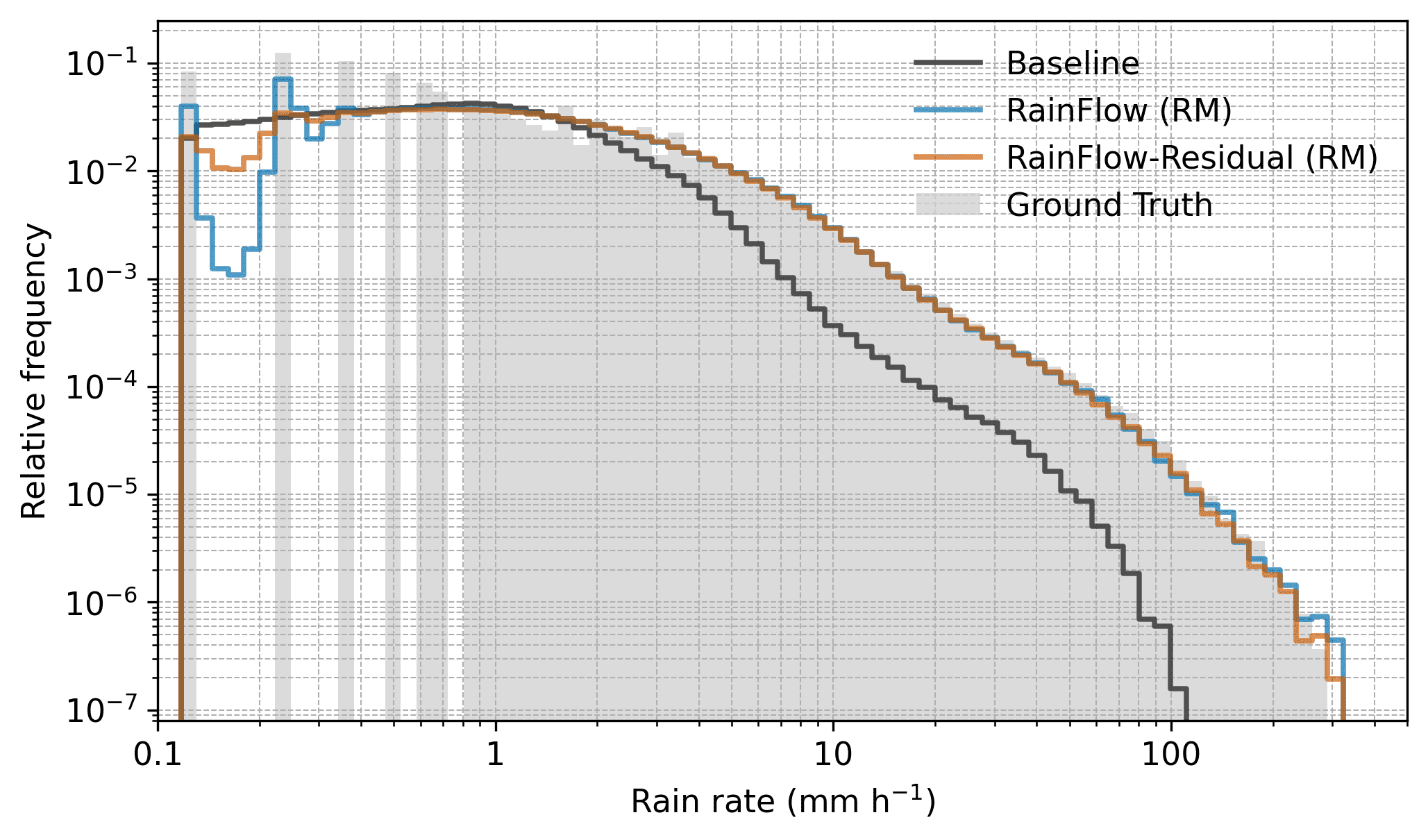}
  \caption{\textbf{Marginal rain distribution (wet pixels).} Histogram of rain rate over all wet test pixels ($RR \geq \SI{0.12}{\milli\metre\per\hour}$) for the observations, the baseline and a randomly drawn member (RM) of each generative model. Members are indistinguishable on this diagnostic (App.~\ref{app:marginal}), so a single draw is representative. The distribution is that of the evaluation set, not the operational climatology (\S\ref{sec:data}).}
  \label{fig:marginal}
\end{figure}

\paragraph{rain rate distribution.} A randomly drawn member of either model reproduces the observed distribution of wet-pixel rain rates across its full range, from the lightest rain up to the heaviest observed intensities, about $\SI{300}{\milli\metre\per\hour}$ (Fig.~\ref{fig:marginal}). The agreement extends to the discrete structure imposed by the radar quantisation at low intensity, which the members largely follow, suggesting that the distribution has been learned in fine detail. It holds for both models and for every member: the $50$ members are indistinguishable through the bulk of the distribution and differ only above about $\SI{100}{\milli\metre\per\hour}$, where pixel counts are small (App.~\ref{app:marginal}). By contrast, the baseline and the ensemble mean (App.~\ref{app:marginal}) stop near $\SI{100}{\milli\metre\per\hour}$. This truncation is well documented for retrievals trained under a pixel-wise loss \citep{ravuri2021skilful, berthomier2023espresso, meignin2026generative, amell2025roa}: the more uncertain the rain rate given the input, as it is for infrared observations \citep{amell2025roa}, the more the mean is pulled towards moderate values, and under weak conditioning the heaviest intensities fall out of its reach. Members are not affected, since each is a sample of the conditional distribution rather than its average, and they reproduce heavy rain at its observed frequency.

\begin{figure}[htbp]
  \centering
  \includegraphics[width=\linewidth]{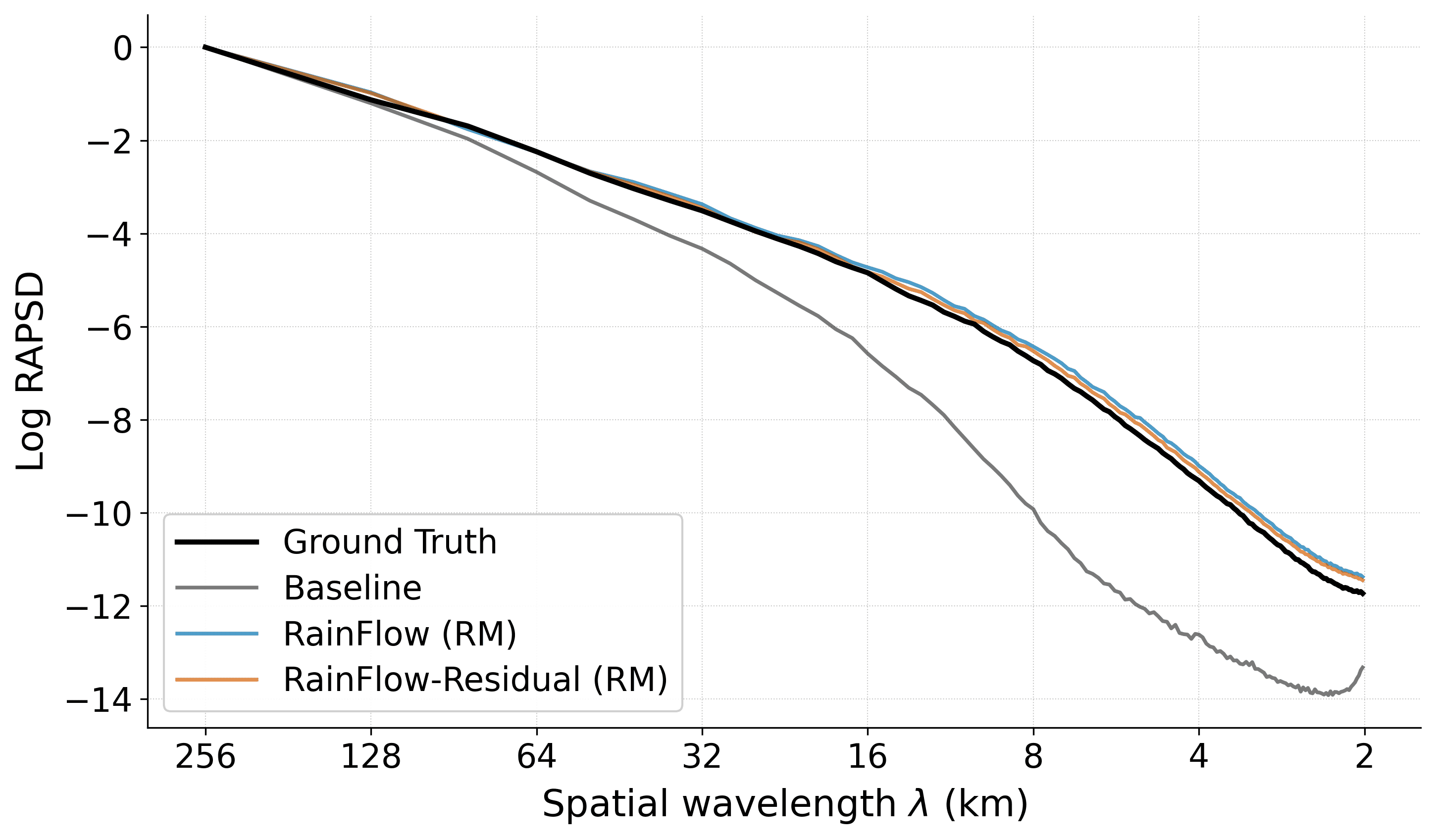}
  \caption{\textbf{Radially averaged power spectra.} Spectra of the observed and predicted rain fields against spatial wavelength, averaged over the $1349$ test scenes, for the baseline and a randomly drawn member (RM) of each generative model. Spectra of the (EM) and of all $50$ members are in App.~\ref{app:psd}.}
  \label{fig:psd}
\end{figure}

\paragraph{Spatial structure.} Neither diagnostic above depends on how rain is arranged in space. We therefore compare the radially averaged power spectra of generated and observed fields, a standard diagnostic of spatial realism for generated precipitation \citep{guilloteau2025difers, meignin2026generative} and atmospheric fields \citep{delefosse2026superresolving}. A randomly drawn member of either model matches the observed spectrum at every resolved scale, from the tile scale down to $\SI{2}{\kilo\metre}$ (Fig.~\ref{fig:psd}), and the two models are again indistinguishable. The baseline, like the ensemble mean (App.~\ref{app:psd}), loses power below roughly $\SI{100}{\kilo\metre}$ and falls increasingly short at smaller scales. This is the spatial counterpart of the truncated tail: plausible fields agree on the large-scale organisation of rain but not on the position of its small-scale features, so averaging them cancels the small-scale variance that each member keeps. 

On all three diagnostics, the generated members are statistically realistic rain fields, a property that the baseline and the ensemble mean, both estimators of the conditional mean, lack under weak conditioning. These diagnostics are, however, invariant to where the rain falls: a realistic field displaced by tens of kilometres would score just as well. Whether the members also put rain in the right location is the question the next section addresses.

\subsection{Location skill}
\label{sec:res-realism}

\begin{figure}[htbp]
  \centering
  \includegraphics[width=\linewidth]{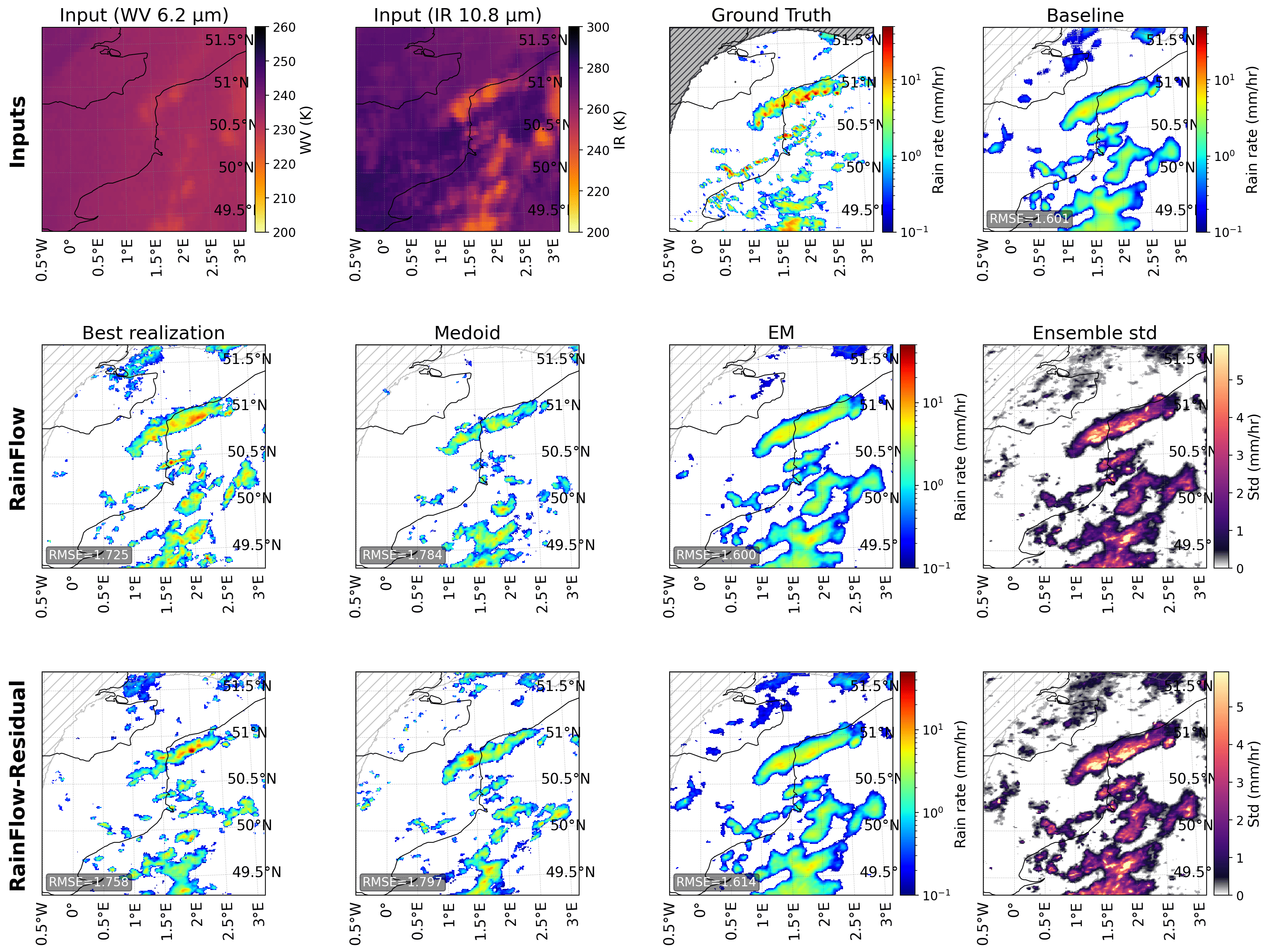}
  \caption{\textbf{Convective case study.} A convective scene over northern France. \emph{Top row:} two of the inputs (WV $6.2$\,\si{\micro\metre}, IR $10.8$\,\si{\micro\metre}), the radar observations and the baseline. \emph{Lower rows}, one per generative model (RainFlow, RainFlow-Residual): the member closest to the radar in RMSE (\emph{best}; selected using the observations), the \emph{medoid} of the ensemble, the EM, and the per-pixel standard deviation over the $50$ members. Rain rate is on a logarithmic scale; RMSE against radar is annotated on each rain panel. Further members are shown in Fig.~\ref{fig:casestudies-appendix}.}
  \label{fig:casestudy}
\end{figure}

\paragraph{Case study.} On a convective scene over northern France (Fig.~\ref{fig:casestudy}), the members reproduce the multi-scale structure of the observed field, with discrete cells of varying intensity separated by dry gaps, while the ensemble mean, like the baseline, merges them into continuous bands: the contrast measured statistically in \S\ref{sec:res-dist} appears here in a single scene. The per-pixel standard deviation is near zero over the dry parts of the domain, where the members agree, and concentrates along the precipitation features, increasing with rain intensity. The ensembles therefore identify where the retrieval is uncertain; the calibration of that spread is evaluated in \S\ref{sec:res-ssr}. A single scene cannot establish how well the generated cells are located relative to the observed ones, and the fractions skill score below quantifies this over the full test year.

\paragraph{Fractions skill score.}
\label{sec:res-fss}
The power spectrum fixes how much variance each scale carries, but not where it lies: cells of the right size and intensity a hundred kilometres from the observed ones give the correct spectrum. We therefore measure location skill with the fractions skill score \citep{roberts2008scale}, which compares the density of rain rate exceedances within a neighbourhood rather than their pixel-by-pixel coincidence. For the ensemble (E), we use its probabilistic form (pFSS; \citealp{necker2024fractions}; App.~\ref{app:evaluation-metrics}), in which the members are converted into a probability of exceedance at each pixel; for the deterministic baseline, it reduces to the classic FSS. A displaced but otherwise correct field is penalised at small neighbourhoods and forgiven at large ones. The quantity of interest is therefore the \emph{skilful scale}: the smallest neighbourhood at which the score reaches the useful-skill level $0.5 + f_o/2$, with $f_o$ the observed exceedance frequency \citep{roberts2008scale}, that is, the finest scale at which the predicted rain can be considered correctly located.

\begin{figure}[htbp]
  \centering
  \includegraphics[width=\linewidth]{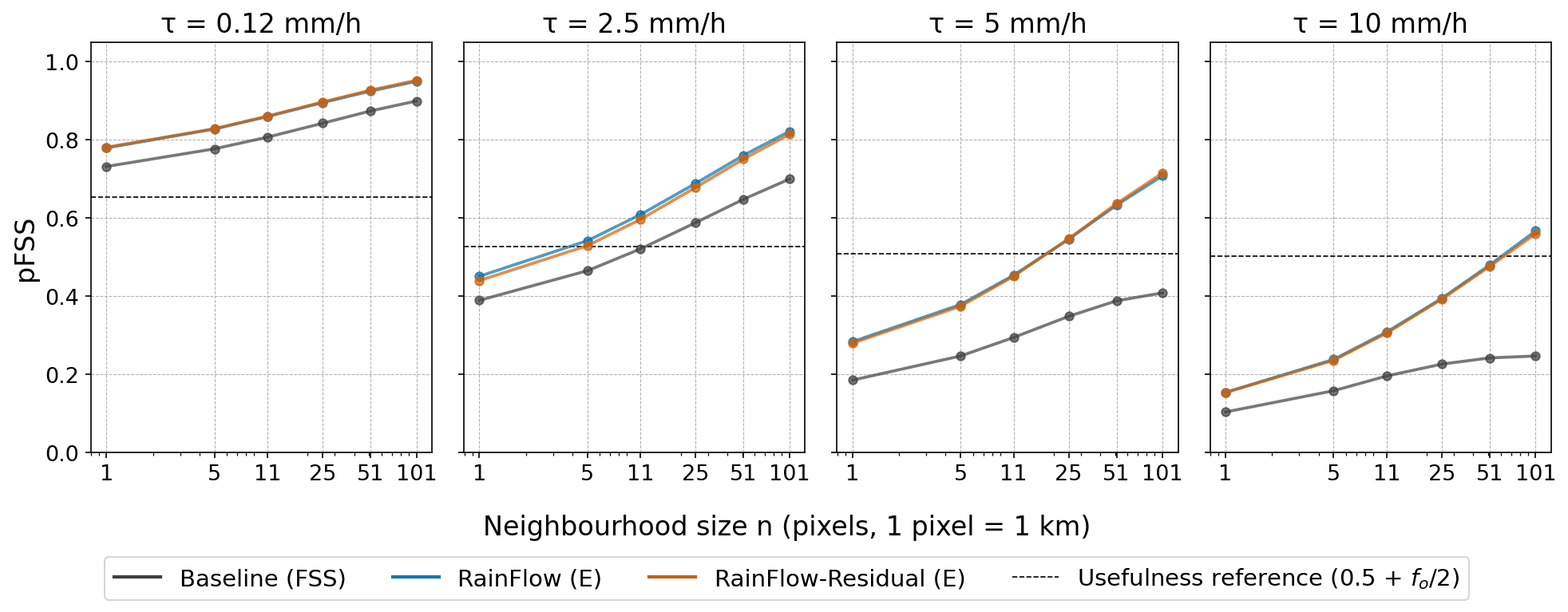}
 \caption{\textbf{Fractions skill score against neighbourhood size.} pFSS of the two generative ensembles (E) and classic FSS of the deterministic baseline, pooled over the $1349$ test scenes, for four rain rate thresholds $\tau$ (App.~\ref{app:evaluation-metrics}). The neighbourhood side $n$ ranges from $1$ to $\SI{101}{\kilo\metre}$. The dashed line is the useful-skill level $0.5 + f_o/2$, with $f_o$ the observed exceedance frequency at that threshold \citep{roberts2008scale}; the neighbourhood at which a curve crosses it is the skilful scale. The FSS of single members, averaged over the $50$ members (AM, avFSS) and for a single draw (RM), is shown in Fig.~\ref{fig:fss-appendix}.}
  \label{fig:pfss}
\end{figure}

Figure~\ref{fig:pfss} shows the curves for four thresholds. At $\tau = \SI{0.12}{\milli\metre\per\hour}$, which separates rain from no rain, every model is skilful from the pixel scale, and the ensembles score above the baseline at every neighbourhood size. Above that threshold, treating the ensemble as a probability of exceedance gives an advantage that grows with intensity. At $\SI{2.5}{\milli\metre\per\hour}$ the ensembles become skilful at about $\SI{5}{\kilo\metre}$, close to the resolution of the infrared input over France, against about $\SI{12}{\kilo\metre}$ for the baseline; at $\SI{5}{\milli\metre\per\hour}$ they become skilful at about $\SI{20}{\kilo\metre}$, while the baseline never does and stays below $0.41$ at the largest neighbourhood tested; and at $\SI{10}{\milli\metre\per\hour}$ they become skilful at about $\SI{60}{\kilo\metre}$, where the baseline remains below $0.25$. The two models are indistinguishable at every threshold and every scale.

The skilful scale nonetheless grows with intensity, from the pixel scale for the presence of rain to several tens of kilometres for heavy rain: infrared observations locate where it rains precisely, but constrain the position of the most intense cells only at coarser scales. The ensemble handles this uncertainty by spreading the probability of heavy rain over the locations where it is plausible, which is what the pFSS rewards.

A large part of this gain comes from pooling the members into a probability. Scored one member at a time and averaged over the $50$ members (AM; Fig.~\ref{fig:fss-appendix}), a member is less well located than the baseline at small scales for the two lowest thresholds, a consequence of its lower detection rate (POD around $0.60$ against $0.88$; \S\ref{sec:res-pixel}), and catches up near $\SI{50}{\kilo\metre}$. At $\SI{5}{\milli\metre\per\hour}$ and above, a single member is ahead of the baseline from about $\SI{10}{\kilo\metre}$ upward, but reaches useful skill only at much coarser scales than the ensemble. The advantage of the ensemble therefore rests on its value as a probability, and the next section evaluates how reliable that probability is: how skilful the ensemble is as a probabilistic prediction, and how well its spread matches the errors it makes.

\clearpage

\subsection{Probabilistic skill and calibration}
\label{sec:res-ssr}

\begin{table}[t]
  \centering
  \small
  \begin{tabular}{lccc}
    \hline
    Model & CRPS & twCRPS, $\tau = 10$ & twCRPS, $\tau = 50$ \\
          & (\si{\milli\metre\per\hour}) & ($10^{-2}$\,\si{\milli\metre\per\hour}) & ($10^{-2}$\,\si{\milli\metre\per\hour}) \\
    \hline
    Baseline & $0.408$ & $3.323$ & $0.343$ \\
    \hline
    RainFlow (E)           & $\mathbf{0.275}$ & $3.009$ & $0.331$ \\
    RainFlow-Residual (E)  & $0.278$ & $\mathbf{3.001}$ & $\mathbf{0.325}$ \\
    \hline
  \end{tabular}
  \caption{\textbf{Probabilistic skill of the full ensemble (E) on the 2019 test set.} CRPS over all pixels, and threshold-weighted CRPS (twCRPS) above $\tau = 10$ and $\SI{50}{\milli\metre\per\hour}$ (App.~\ref{app:evaluation-metrics}); lower is better. The baseline, a prediction without spread for which both scores reduce to an absolute error, is given for reference. Best per column in bold.}
  \label{tab:crps}
\end{table}

\paragraph{Probabilistic skill.} We score the full ensemble (E) with the continuous ranked probability score (CRPS), a proper score that generalises the absolute error to a predictive distribution: its expected value is minimised by the true conditional distribution, and so penalises an ensemble that is too narrow as well as one that is too wide. To focus on heavy rain, we also use its threshold-weighted form (twCRPS; \citealp{gneiting2011comparing}), which scores only the part of the predictive distribution above a threshold. Unlike a CRPS computed on pixels binned by observed intensity, which conditions on the outcome and rewards models that overpredict heavy rain everywhere \citep{lerch2017forecaster}, the twCRPS remains proper.

The two models are nearly indistinguishable (Table~\ref{tab:crps}). Over all pixels, RainFlow scores marginally better ($0.275$ against $0.278$), whereas above $\SI{10}{\milli\metre\per\hour}$ and $\SI{50}{\milli\metre\per\hour}$ RainFlow-Residual does. These differences are too small to establish a ranking between the two formulations (App.~\ref{app:tsampling}). At most, they suggest a slight edge of the residual formulation for heavy rain, which would be consistent with its design and section~\ref{sec:res-cond}, in which the baseline supplies the amplitude of the field and the flow generates only the variability around it.

The clearest difference between the two models lies elsewhere: their sensitivity to a training choice usually left at its default, the density from which the training time is sampled (App.~\ref{app:tsampling}). Between the uniform and power-law densities, the CRPS of RainFlow changes by $3.6\%$, and by up to $12\%$ when the logit-normal density is included, more than the gap between the two models. Over the same two densities, the CRPS of RainFlow-Residual changes by only $0.4\%$, about nine times less, and by $1.4\%$ across all three. The mechanism is the endpoint argument of \S\ref{sec:time-sampling}: the baseline supplies the amplitude of the field, so the flow never has to recover it from the infrared conditioning alone at the start of the trajectory, where the sampling density matters most. The residual formulation is therefore less a source of skill than a source of robustness: it makes the skill of the model nearly insensitive to a training choice that otherwise weighs more than the formulation itself, although the spread of the ensemble still depends on it.

\begin{table}[t]
  \centering
  \small
  \begin{tabular}{lccc}
    \hline
    Model & Spread (\si{\milli\metre\per\hour}) & RMSE (\si{\milli\metre\per\hour}) & SSR \\
    \hline
    RainFlow (E)          & $1.326$ & $1.495$ & $\mathbf{0.887}$ \\
    RainFlow-Residual (E) & $1.312$ & $\mathbf{1.487}$ & $0.882$ \\
    \hline
  \end{tabular}
\caption{\textbf{Spread--skill ratio (SSR) on the 2019 test set}, over all test pixels. Spread is the square root of the mean ensemble variance over the $M = 50$ members, and skill the RMSE of the ensemble mean (Table~\ref{tab:em-pixel}); both are defined in App.~\ref{app:evaluation-metrics}. A calibrated ensemble scores $1$; below $1$ is underdispersed.}
  \label{tab:ssr}
\end{table}

\begin{figure}[t]
  \centering
  \includegraphics[width=\linewidth]{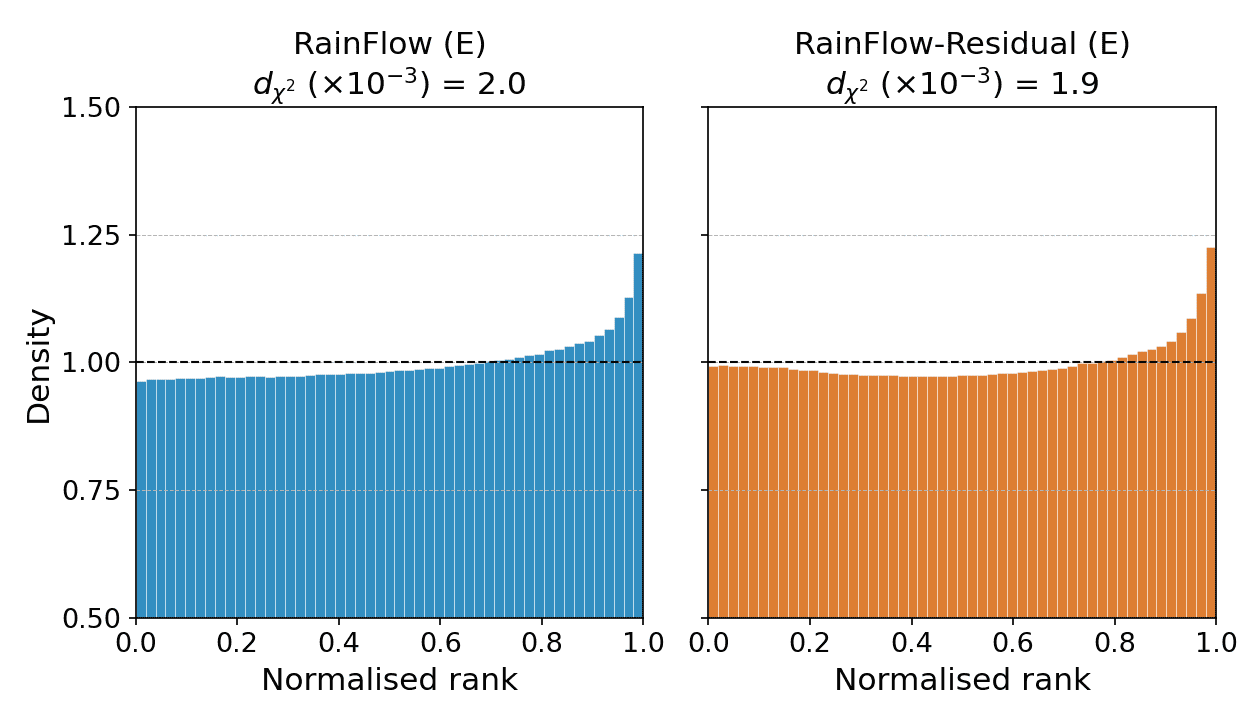}
  \caption{\textbf{rank histograms of the full ensemble (E).} Randomised rank of the radar observation among the 50 members of RainFlow (left) and RainFlow-Residual (right), over all valid pixels of the 1349 test scenes, with one bin per rank (51 bins; App.~\ref{app:evaluation-metrics}). The value $d_{\chi^2}$ in each panel title is the mean squared departure of the bars from the dashed line; it is zero for a perfectly flat histogram. The dashed line is the flat histogram of a calibrated ensemble; an excess in the highest bins means that observations exceed all members too often. Pixels where the observation and all members are dry receive random ranks and are spread evenly over the bins.}
  \label{fig:rank}
\end{figure}

\paragraph{Calibration.} The spread--skill ratio compares the ensemble spread with the RMSE of the ensemble mean; it equals one for a calibrated ensemble and falls below one for an underdispersed one \citep{fortin2014should}. Both ensembles are mildly underdispersed, with ratios of $0.887$ for RainFlow and $0.882$ for RainFlow-Residual (Table~\ref{tab:ssr}): their spread is of the right order of magnitude, covering most of the error of the ensemble mean, but is slightly too narrow. The rank histogram of the probability integral transform (PIT; App.~\ref{app:evaluation-metrics}) shows where in the distribution this shortfall lies (Fig.~\ref{fig:rank}). For both models, it is nearly flat except in the highest ranks: observations exceed all members slightly more often than a calibrated ensemble would allow, whereas the opposite case is not in excess. The underdispersion is thus one-sided, the ensembles missing the highest values rather than being too narrow on both sides. RainFlow also tilts slightly towards high ranks, the signature of a small low bias. Both diagnostics are pooled over all pixels, however, so neither shows where the ensembles fall short; the next section examines how the generated rain depends on the cloud-top temperature. 

\subsection{Conditioning on brightness temperature}
\label{sec:res-cond}

The diagnostics so far compare the generated fields with the observations, but none tests how the generated rain depends on the conditioning. A model that produced realistic rain fields regardless of the infrared input would pass the tests of \S\ref{sec:res-dist}, and the fractions skill score of \S\ref{sec:res-realism} tests where rain falls, not whether its amount and variability follow the input. This dependence is, however, what a retrieval is meant to learn, and under weak conditioning it must be learned from a weak signal. We therefore examine the rain rate as a function of the brightness temperature of the $10.8$\,\si{\micro\metre} window channel, $\mathrm{IR}_{10.8}$. The infrared channels are strongly correlated, and this channel, the one most directly tied to cloud-top height, is the basis of operational infrared precipitation products, some of which use it alone \citep{janowiak2001, hong2004ccs}. The models remain conditioned on all seven channels; only the diagnostic is restricted to this one.

Since $\mathrm{IR}_{10.8}$ is one of the model inputs, we can state in advance how each model should behave. The baseline estimates the mean rain rate for each input. Its individual estimates may differ widely from the observed values, but when they are averaged over all pixels with the same cloud-top temperature, the result should agree with the observed mean at that temperature, even when the conditioning is weak. The standard deviation behaves differently. Rain rates vary among pixels with the same cloud-top temperature for two reasons: the other channels and the surrounding scene differ from one pixel to the next, and the cloud top does not fully determine the rain beneath it. The baseline can represent the first source of variability but not the second, since it returns a single value for each input. Its conditional standard deviation should therefore fall below the observed one, whereas the generated members, which sample the range of rain rates compatible with each input, should reproduce it. The conditional mean is thus a test for all models, and the conditional standard deviation tests whether the members supply the variability that the input leaves undetermined. For the ensembles, the $50$ members are pooled (50M) before these statistics are computed. All pixels are retained: selecting them on the observed rain rate would let the observations choose the sample and mix intensity errors with the location errors of \S\ref{sec:res-realism} (App.~\ref{app:tb}).

\begin{figure}[htbp]
  \centering
  \includegraphics[width=0.49\linewidth]{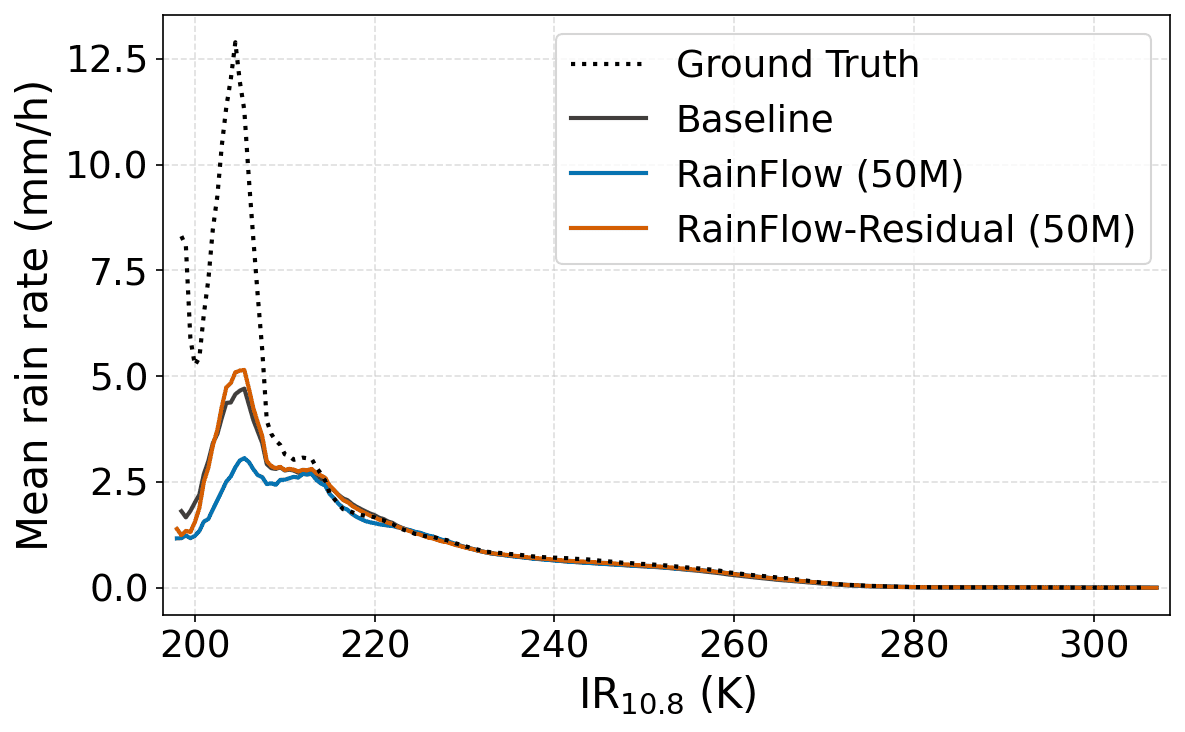}\hfill
  \includegraphics[width=0.49\linewidth]{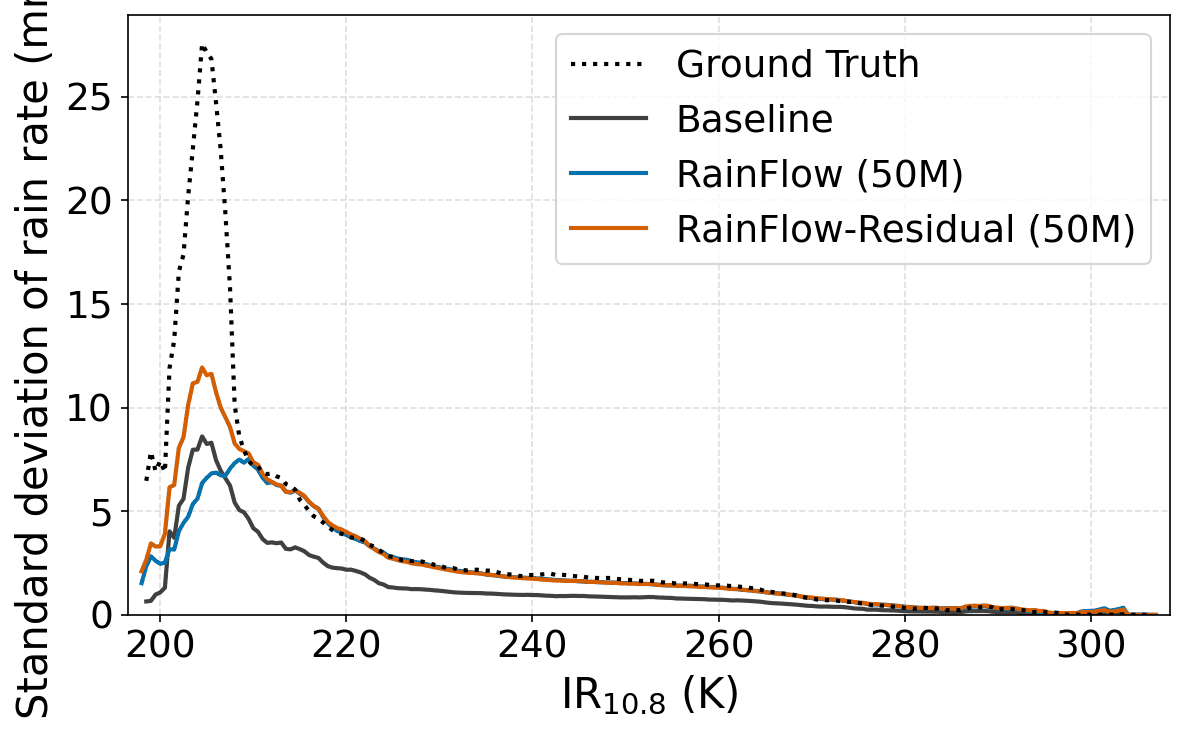}
  \caption{\textbf{Conditional moments of rain rate given brightness temperature.} Mean (\emph{left}) and standard deviation (\emph{right}) of $RR$ in each $\mathrm{IR}_{10.8}$ bin, over all pixels of the $1349$ test scenes, with the $50$ members pooled (50M) for the ensembles, against the observed values (dotted). A best estimate such as the baseline should match the observed mean but fall below the observed standard deviation. Curves restricted to observed wet pixels are in App.~\ref{app:tb}.}
  \label{fig:condtb}
\end{figure}

Over most of the brightness temperature range, the results follow these expectations (Fig.~\ref{fig:condtb}). Above roughly $\SI{215}{\kelvin}$, all three models follow the observed conditional mean closely. The conditional standard deviation separates them: both ensembles follow the observed curve, whereas the baseline lies below it almost everywhere, by roughly a factor of two in the bulk. This gap is the variability that a single best estimate cannot represent, and the members supply it: for all but the coldest cloud tops, the generated rain depends on the brightness temperature as the observed rain does, in both amount and variability.

Below $\SI{215}{\kelvin}$, all three models underestimate both moments. Near $\SI{205}{\kelvin}$, the observed mean and standard deviation reach $12.8$ and $\SI{27.6}{\milli\metre\per\hour}$, against $5.2$ and $11.9$ for RainFlow-Residual, $4.7$ and $8.6$ for the baseline, and $3.1$ and $\SI{6.9}{\milli\metre\per\hour}$ for RainFlow. Each ensemble falls short on both moments by about the same factor, about $2.4$ for RainFlow-Residual and $4$ for RainFlow: the generated distribution has roughly the right shape but too low an amplitude. Weak conditioning alone cannot explain this, since even the baseline should reproduce the observed mean, and neither can the location of the rain: parallax and the offset between cloud top and surface rain are also present in the training data, so a model that had learned this relation would reproduce the observed curves despite them. The deficit is common to three models trained with different objectives, and its most likely cause is the scarcity of such cold tops, and of the heavy rain beneath them, in the training data.

\begin{figure}[htbp]
  \centering
  \includegraphics[width=\linewidth]{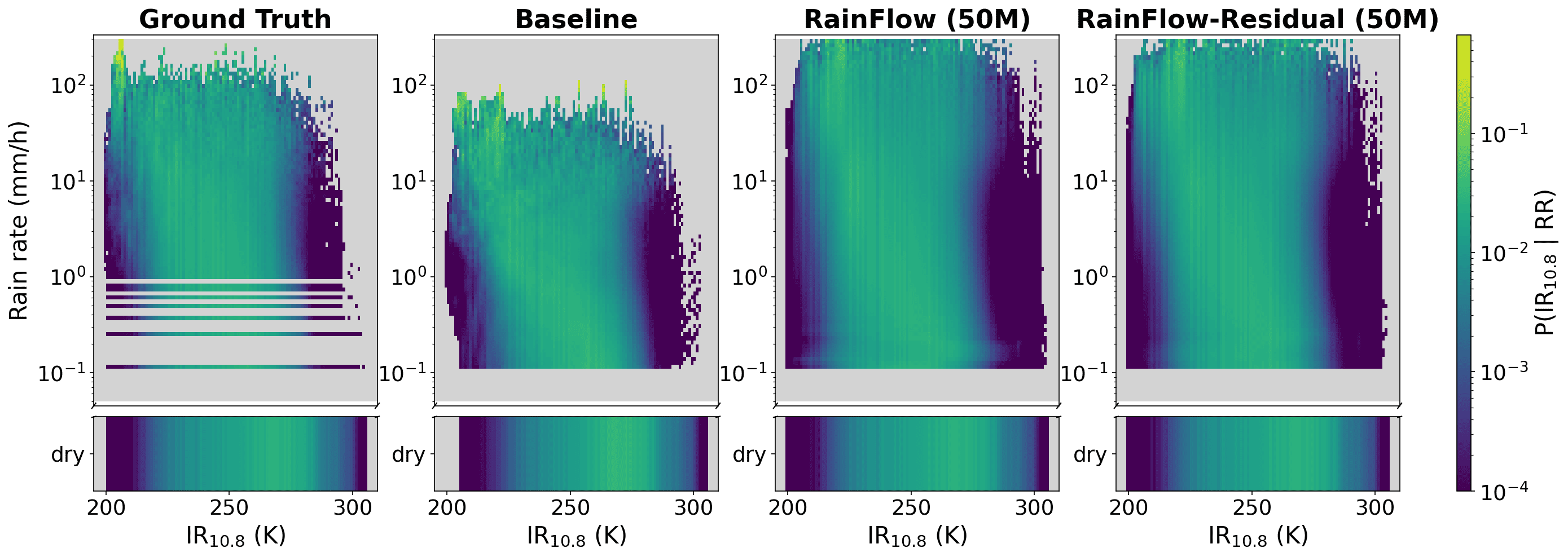}
  \caption{\textbf{Distribution of cloud-top temperature for each rain rate, $P(\mathrm{IR}_{10.8} \mid RR)$.} Each row is normalised and gives the distribution of $\mathrm{IR}_{10.8}$ over the pixels where the observations, the baseline or the $50$ pooled members (50M) of each generative model have that rain rate, over the $1349$ test scenes; the lower strip is the dry class. The outer edges of each panel extend with the number of pixels, which is fifty times larger for the pooled members.}
  \label{fig:ptbr}
\end{figure}

Figure~\ref{fig:ptbr} takes the opposite view: each row shows how the pixels with a given rain rate, observed or generated, are distributed over cloud-top temperature. In the observations, the heaviest rain concentrates in the top-left corner of the panel, under the coldest tops, although parallax and the offset between cloud top and surface rain leave some heavy rain under warm co-located tops. The baseline does not produce rain that heavy (\S\ref{sec:res-dist}). The generative models do, but spread it almost evenly over all cloud-top temperatures, so the top-left corner holds a much smaller share of their heavy rain. RainFlow-Residual is closer to the observations than RainFlow in this respect. The two figures thus show the same gap from opposite sides: under the coldest tops, the generated rain is too weak, and above heavy generated rain, the tops are too rarely the coldest.

\section{Discussion}
\label{sec:discussion}

Over most of the brightness temperature range, the ensembles reproduce both the mean and the spread of the rain rate given the infrared input (\S\ref{sec:res-cond}). The main exception is deep convection under the coldest cloud tops, below about $\SI{215}{\kelvin}$, where all three models underestimate both quantities. Such tops are rare over France, where precipitation is predominantly of mid-latitude type, and under weak conditioning these few examples are likely insufficient to learn the relationship precisely. The heavy tail of the marginal distribution is nevertheless well reproduced (\S\ref{sec:res-dist}): every heavy-rain example contributes to it, whereas its dependence on the coldest temperatures can only be learned from the subset observed at those temperatures. The baseline shows the same deficit, which suggests that the limitation stems from the training data rather than from the generative approach, and that this regime would mainly benefit from additional deep-convective examples.

Outside this regime, the two models perform similarly, and the difference between direct and residual generation is smaller than the effect of the density from which training times are drawn. With the density of \S\ref{sec:time-sampling}, which favours the noise end of the trajectory, RainFlow reaches the CRPS of RainFlow-Residual, with a 1.2\% lead; with the uniform and logit-normal densities, it falls behind by 2.2\% and 9.2\%, respectively (\S\ref{sec:res-ssr}, App.~\ref{app:tsampling}). The endpoint argument of \S\ref{sec:time-sampling} accounts for this asymmetry: near the noise end, RainFlow must learn the conditional mean from the infrared conditioning alone, whereas RainFlow-Residual receives it from the baseline prediction. In this setting, the main benefit of the residual formulation is therefore its robustness to the time density, together with a small advantage for heavy rain.

This observation is relevant beyond the present problem. Residual formulations are widely used in generative downscaling, retrieval and forecasting \citep{mardani2025residual, guilloteau2025difers, couairon2026archesweathergen, delefosse2026superresolving}, generally without a comparison to direct generation: \citet{mardani2025residual} motivate theirs by the poor samples of a directly trained diffusion model, which they attribute to the high noise levels the direct target requires, precisely the regime that our power-law density emphasises, and the two flow-matching studies train with the logit-normal density \citep{couairon2026archesweathergen, delefosse2026superresolving}, the least favourable to direct generation in our comparison. \citet{rampal2026intercomparison} compare both formulations for precipitation downscaling and report mixed results across model families. None of these studies varies the time density, and our results suggest that it should be controlled for when the two formulations are compared.

The weak conditioning is also partly instrumental. Beyond the irreducible gap between cloud-top temperature and surface rain, the SEVIRI channels are acquired at roughly $\SI{6}{\kilo\metre}$ and upsampled to the $\SI{1}{\kilo\metre}$ radar grid, and the two contributions cannot be separated with this data. Spectral agreement below the input resolution therefore reflects realistic texture rather than retrieval skill, and the location of that texture is constrained only at the scales the input resolves (\S\ref{sec:res-realism}). Separating the two would require a coarser reference or a higher-resolution instrument, such as the imager of Meteosat Third Generation.

These results should be read within the scope of the experiment. The study compares models on a common evaluation set rather than validating an operational product: the set is deliberately oversampled toward precipitating scenes, so aggregate frequencies such as the dry fraction describe that set rather than the operational climatology, and validation under natural sampling is a separate exercise. The results are for a single mid-latitude domain, a single sensor and a single radar reference.

\section{Conclusion}
\label{sec:conclusion}

This study formulates precipitation retrieval from infrared observations alone as conditional generation and introduces RainFlow, a flow-matching model that generates ensembles of rain fields for this problem, in two variants: one generating rain fields directly, the other generating the variability around a deterministic prediction. Evaluated against radar over a full year, both improve substantially on a baseline that estimates the conditional mean. Individual members reproduce the observed distribution of rain rates, including its heavy tail, as well as the dry fraction and the power spectrum, none of which the baseline reproduces. The ensembles also represent the variability of the rain rate under cloud tops of a given temperature over most of the brightness temperature range, where the baseline underestimates it throughout. The ensemble mean remains as accurate an estimator of the conditional mean as the baseline, so these improvements do not come at the expense of accuracy. Used as probabilities of exceedance, the ensembles locate rain with higher skill than the baseline, with a gain that grows with rain intensity, and their spread covers most of the error of their mean.

The comparison between the two variants also informs the design of such models. Direct and residual generation reach similar skill when training times are drawn from a density that favours the noise end of the trajectory, where the network learns the conditional mean from the infrared input. For direct generation, this density has a larger effect than the choice of formulation, whereas the skill of the residual variant is nearly insensitive to it and retains a small advantage for heavy rain. Comparisons between the two formulations, which have so far not considered the time density, should therefore account for it.

Generative retrievals are mostly evaluated on their outputs alone, against the observations. Evaluating the generated rain as a function of the conditioning adds a complementary diagnostic: it tests whether the models reproduce how rain depends on the infrared conditioning, which is precisely what weak conditioning makes difficult to learn. Here, it reveals a limitation that the other diagnostics do not show: although the heavy tail of the marginal distribution is well reproduced, all three models, including the baseline, underestimate both the amount of rain and its variability under the coldest cloud tops, the regime least represented in a mid-latitude dataset. We therefore recommend including such conditional diagnostics in the evaluation of generative retrievals, since they assess what these models are meant to learn: the distribution of rain given the conditioning.

Several directions follow from these results. The models currently use a single infrared image, whereas the temporal evolution of a cloud top carries additional information: a rapidly cooling top indicates a developing convective core, while an older anvil at the same temperature produces little rain. Including the temporal dynamics of the infrared observations could therefore strengthen the conditioning where it is weakest. The same regime would also benefit from additional deep-convective examples, from regions with tropical convection or through oversampling of cold-top situations during training. On the modelling side, the numerical integration, which requires many network evaluations per member, could be reduced to one or a few evaluations, either by straightening the learned trajectories with reflow \citep{liu2022flowstraightfastlearning, lee2024improving} or with one-step generators such as consistency models \citep{song2023consistency}, shortcut models \citep{frans2025shortcut} or MeanFlow \citep{geng2025meanflow}, substantially reducing the cost of generating large ensembles. Faster sampling would also make it practical to select checkpoints on the CRPS and to use it as a training objective; as a proper score, the CRPS penalises underdispersion and could thus help correct the mild underdispersion observed here. Finally, the higher resolution of the Meteosat Third Generation imager would help separate the physical and instrumental contributions to the weak conditioning.

\clearpage

\acknowledgments

This work was supported by the \textit{Terre-Océan-Surfaces Continentales-Atmosphère} program of the Centre National d'Etudes Spatiales (CNES) and the \textit{Programme National de Télédétection}. Computing and storage resources were provided by GENCI at IDRIS through Grant 2024-AD011012790R4 on the Jean Zay supercomputer's V100 partition.  

Additional support was provided by the France 2030 project \textit{Académie Spatiale d'Île-de-France} (\url{https://academiespatiale.fr/}), managed by the National Research Agency under reference ANR-23-CMAS-0041.

\datastatement
The training data cannot be redistributed under the licences of its two sources, but both are openly available. The MSG/SEVIRI Level 1.5 infrared brightness temperatures are distributed by EUMETSAT through the EUMETSAT Data Centre (\url{https://data.eumetsat.int/product/EO:EUM:DAT:MSG:HRSEVIRI}), and the M\'et\'eo-France $5$-minute precipitation radar mosaics used as reference are available from Aeris (\url{https://doi.org/10.25326/357}). Construction of the matchup set from these two sources follows \citet{meignin2026generative}.

All implementation details and the full training code is available at \url{https://github.com/mmeignin/RainFlow}.

\clearpage

\appendix[A]

\appendixtitle{Evaluation metrics}
\applabel{A}{app:evaluation-metrics}

We define here the metrics used to evaluate the models in Section~\ref{sec:results}. For single-member models, the second terms of the CRPS and twCRPS are dropped, so that both reduce to absolute errors, and the SSR and rank histogram are not computed.

\paragraph{CRPS.}
We evaluate the predictive distribution at each grid point with the continuous ranked probability score \citep{gneiting2007strictly}, using the fair estimator of \citet{zamo2018estimation}. For an ensemble of $M$ members $\hat{y}_1, \dots, \hat{y}_M$ and observation $y$:
\begin{equation}
  \mathrm{CRPS}
  = \frac{1}{M} \sum_{n=1}^{M} \lvert \hat{y}_n - y \rvert
  - \frac{1}{2M(M-1)} \sum_{n=1}^{M} \sum_{n'=1}^{M} \lvert \hat{y}_n - \hat{y}_{n'} \rvert .
\end{equation}
Lower is better.

\paragraph{Threshold-weighted CRPS.}
To evaluate heavy precipitation, we use the threshold-weighted CRPS \citep{gneiting2011comparing}, which restricts the CRPS integral to values above a threshold $\tau$ through the weight $w(z) = \mathbb{1}\{z \ge \tau\}$. As in \citet{wessel2025improving}, it is computed as the CRPS of the censored ensemble against the censored observation, with $v_\tau(z) = \max(z, \tau)$:
\begin{equation}
  \mathrm{twCRPS}_\tau
  = \frac{1}{M} \sum_{n=1}^{M} \bigl\lvert v_\tau(\hat{y}_n) - v_\tau(y) \bigr\rvert
  - \frac{1}{2M(M-1)} \sum_{n=1}^{M} \sum_{n'=1}^{M} \bigl\lvert v_\tau(\hat{y}_n) - v_\tau(\hat{y}_{n'}) \bigr\rvert .
\end{equation}
Since $v_\tau$ maps every value below $\tau$ to $\tau$, the score does not distinguish between values below the threshold: it depends only on the probability the ensemble assigns to exceeding $\tau$ and on how that probability is distributed above $\tau$. We use fixed thresholds $\tau = 10$ and $50~\mathrm{mm\,h^{-1}}$, so that scores are comparable across models. Precipitation being non-negative, the twCRPS with $\tau = 0$ is the CRPS.

\paragraph{SSR.}
The spread--skill ratio compares the spread of the ensemble with the error of its mean. At each grid point, the ensemble mean and the unbiased ensemble variance are
\begin{equation}
  \hat{y}_{\mathrm{EM}} = \frac{1}{M} \sum_{n=1}^{M} \hat{y}_n ,
  \qquad
  s^2 = \frac{1}{M-1} \sum_{n=1}^{M} \big( \hat{y}_n - \hat{y}_{\mathrm{EM}} \big)^2 .
\end{equation}
The variance is corrected for the finite ensemble size by the factor $(M+1)/M$, the spread is the square root of the mean corrected variance, the skill is the RMSE of the ensemble mean, and the spread--skill ratio is their quotient:
\begin{equation}
  \mathrm{Spread} = \sqrt{\Big\langle \tfrac{M+1}{M}\, s^2 \Big\rangle} ,
  \qquad
  \mathrm{RMSE} = \sqrt{\big\langle (\hat{y}_{\mathrm{EM}} - y)^2 \big\rangle} ,
  \qquad
  \mathrm{SSR} = \frac{\mathrm{Spread}}{\mathrm{RMSE}} ,
\end{equation}
where $\langle \cdot \rangle$ denotes the average over all grid points of the test samples with a valid observation. Variances and squared errors are thus averaged before taking the square roots. With this correction, $\mathrm{SSR} = 1$ for a statistically consistent ensemble \citep{fortin2014should}; values below (above) 1 indicate an under- (over-)dispersive ensemble.

\paragraph{RAPSD.}
To compare the spatial structure of generated and observed fields, we use the radially averaged power spectral density \citep{ruzanski2011scale}, following pysteps \citep{pulkkinen2019pysteps}. For a tile of $L \times L$ pixels with grid spacing $\Delta x$, let $\hat{X}(\boldsymbol{\kappa})$ be its discrete Fourier transform at the integer wavevector $\boldsymbol{\kappa} = (\kappa_x, \kappa_y)$, $\kappa_x, \kappa_y \in \{-L/2, \dots, L/2 - 1\}$. Let $\mathcal{R}_k$ be the set of wavevectors with $\operatorname{round}\lVert \boldsymbol{\kappa} \rVert = k$, and $N_k$ their number. Then
\begin{equation}
  E(k) = \frac{1}{N_k} \sum_{\boldsymbol{\kappa} \in \mathcal{R}_k}
         \bigl\lvert \hat{X}(\boldsymbol{\kappa}) \bigr\rvert^2 ,
  \qquad k = 1, \dots, L/2 - 1 ,
\end{equation}
where wavenumber $k$ corresponds to a wavelength of $L\,\Delta x / k$. Spectra are shown as $\ln E(k) - \ln E(1)$, i.e.\ relative to the largest resolved scale.

\paragraph{Rank histogram.}
The rank histogram \citep{talagrand1997evaluation, hamill2001interpretation}, the ensemble counterpart of the probability integral transform (PIT) histogram \citep{gneiting2007probabilistic}, assesses the calibration of the ensemble. At each grid point, the observation $y$ is ranked among the $M+1$ values formed by the $M$ members and $y$ itself. If the observation is statistically indistinguishable from a member, its rank is uniform on $\{0, \dots, M\}$. Ties occur at every dry grid point, where members and observation are exactly zero, and the rank is then drawn uniformly among the tied positions. With $M_<$ and $M_=$ the numbers of members below and equal to $y$, we compute the unified PIT \citep{vogel2018skill}
\begin{equation}
  u = \frac{M_< + V\,(M_= + 1)}{M + 1}, \qquad V \sim \mathcal{U}(0, 1).
\end{equation}
On $M+1$ equal-width bins, the bin index of $u$ is $\lfloor (M+1)\,u \rfloor = M_< + \lfloor V\,(M_= + 1) \rfloor$, which is the rank with ties broken at random. The histogram of $u$ over these bins is therefore the rank histogram, and it is expected to be flat for a calibrated ensemble. A U shape indicates underdispersion, a hump over-dispersion, and an asymmetric histogram a bias. A flat histogram is necessary but not sufficient for calibration, and a U shape can also come from conditional biases \citep{hamill2001interpretation}.

\paragraph{Ensemble-mean metrics.}
We also evaluate the ensemble mean $\hat{y}_{\mathrm{EM}}$ as a point prediction. For single-member models, $\hat{y}_{\mathrm{EM}}$ is the model output itself. With $N$ the total number of valid grid points over all test samples, and $\hat{y}_{\mathrm{EM},i}$ and $y_i$ the ensemble mean and the observation at the $i$-th of these points,
\begin{equation}
  \mathrm{RMSE} = \sqrt{\frac{1}{N} \sum_{i=1}^{N} \bigl( \hat{y}_{\mathrm{EM},i} - y_i \bigr)^2} ,
  \qquad
  \mathrm{MAE} = \frac{1}{N} \sum_{i=1}^{N} \bigl\lvert \hat{y}_{\mathrm{EM},i} - y_i \bigr\rvert .
\end{equation}
For rain detection, grid point $i$ is predicted wet when $\hat{y}_{\mathrm{EM},i} \ge 0.12~\mathrm{mm\,h^{-1}}$ and observed wet when $y_i \ge 0.12~\mathrm{mm\,h^{-1}}$. Over the $N$ valid grid points, we count the true positives $\mathrm{TP}$ (predicted and observed wet), true negatives $\mathrm{TN}$ (predicted and observed dry), false positives $\mathrm{FP}$ (predicted wet but observed dry) and false negatives $\mathrm{FN}$ (predicted dry but observed wet), so that $\mathrm{TP} + \mathrm{TN} + \mathrm{FP} + \mathrm{FN} = N$. From these, we compute the critical success index, the probability of detection and the false alarm ratio:
\begin{equation}
  \mathrm{CSI} = \frac{\mathrm{TP}}{\mathrm{TP} + \mathrm{FP} + \mathrm{FN}} , \qquad
  \mathrm{POD} = \frac{\mathrm{TP}}{\mathrm{TP} + \mathrm{FN}} , \qquad
  \mathrm{FAR} = \frac{\mathrm{FP}}{\mathrm{TP} + \mathrm{FP}} .
\end{equation}
Higher CSI and POD and lower FAR are better. All scores are computed from errors and counts pooled over the valid grid points of all test samples, not averaged per sample.

\paragraph{Single-member metrics.}
To evaluate individual realisations rather than their average, we compute the same scores for each member $n = 1, \dots, M$ on its own, replacing $\hat{y}_{\mathrm{EM},i}$ by $\hat{y}_{n,i}$:
\begin{equation}
  \mathrm{RMSE}_n = \sqrt{\frac{1}{N} \sum_{i=1}^{N} \bigl( \hat{y}_{n,i} - y_i \bigr)^2} ,
  \qquad
  \mathrm{MAE}_n = \frac{1}{N} \sum_{i=1}^{N} \bigl\lvert \hat{y}_{n,i} - y_i \bigr\rvert ,
\end{equation}
and likewise $\mathrm{CSI}_n$, $\mathrm{POD}_n$ and $\mathrm{FAR}_n$ from the contingency counts of member $n$. Each score is pooled over the valid grid points of all test samples, and we report its mean and standard deviation over the $M$ members. For single-member models these coincide with the ensemble-mean metrics. By the triangle inequality, the member-averaged RMSE and MAE are never lower than those of the ensemble mean, whereas no such ordering holds for CSI, POD and FAR.

\paragraph{Fractions skill score.}
For a threshold $\tau$, the binary probabilities $\mathrm{BP}_{ij,o}$ and $\mathrm{BP}_{ijn,f}$ equal 1 if the observation $y$ and the member $\hat{y}_n$, respectively, reach $\tau$ at the $j$-th grid point of the $g \times g$ neighbourhood centred on grid point $i$, and 0 otherwise. The observed and forecast neighbourhood probabilities are
\begin{equation}
  \mathrm{NP}_{i,o} = \frac{1}{g^2} \sum_{j=1}^{g^2} \mathrm{BP}_{ij,o} , \qquad \mathrm{NP}_{in,f} = \frac{1}{g^2} \sum_{j=1}^{g^2} \mathrm{BP}_{ijn,f} .
\end{equation}
For a forecast neighbourhood probability $\mathrm{NP}_{i,f}$, the fractions Brier score (FBS), its worst value and the fractions skill score \citep{roberts2008scale} are
\begin{equation}
  \mathrm{FBS} = \sum_{i} \bigl( \mathrm{NP}_{i,f} - \mathrm{NP}_{i,o} \bigr)^2 , \qquad \mathrm{FBS}_{\mathrm{worst}} = \sum_{i} \mathrm{NP}_{i,f}^2 + \sum_{i} \mathrm{NP}_{i,o}^2 , \qquad \mathrm{FSS} = 1 - \frac{\mathrm{FBS}}{\mathrm{FBS}_{\mathrm{worst}}} ,
\end{equation}
which ranges from 0 to 1, with 1 a perfect forecast. With $\mathrm{FSS}_n$ the score of member $n$, obtained with $\mathrm{NP}_{i,f} = \mathrm{NP}_{in,f}$, we compute three variants following \citet{necker2024fractions}: the FSS of a single member, $\mathrm{FSS}_1$; the ensemble-averaged FSS (avFSS), the mean of $\mathrm{FSS}_n$ over members; and the probabilistic FSS (pFSS), obtained by replacing $\mathrm{NP}_{i,f}$ with the neighbourhood ensemble probability (NEP):
\begin{equation}
  \mathrm{avFSS} = \frac{1}{M} \sum_{n=1}^{M} \mathrm{FSS}_n , \qquad \mathrm{NEP}_{i,f} = \frac{1}{M} \sum_{n=1}^{M} \mathrm{NP}_{in,f} = \frac{1}{M g^2} \sum_{j=1}^{g^2} \sum_{n=1}^{M} \mathrm{BP}_{ijn,f} .
\end{equation}
For single-member models, the three variants reduce to the classic FSS. We use $g = 1, 5, 11, 25, 51$ and $101$ grid points and $\tau = 0.12, 2.5, 5$ and $10~\mathrm{mm\,h^{-1}}$.

\clearpage

\appendix[B]

\appendixtitle{Pixel-wise scores of single members}
\applabel{B}{app:member-pixel}

Table~\ref{tab:pixel} reports the pixel-wise scores of Table~\ref{tab:em-pixel} for single members, averaged over the $50$ members (AM). A member is a draw from $p(RR \mid c)$, not an estimate of its expectation: if members were drawn from the true conditional distribution, their mean squared error would exceed that of the conditional mean by the conditional variance. The higher RMSE of the members ($1.98$ and $1.97$, against $1.48$ for the baseline) is this expected penalty rather than a deficiency of the models.

The categorical scores show the same contrast from the other side. A member commits to one location of the rain cells, separated by dry gaps, so it detects less of the observed rain than the smooth baseline ($\mathrm{POD} = 0.60$ and $0.61$ against $0.88$) but raises fewer false alarms ($\mathrm{FAR} = 0.35$ against $0.40$). The ensemble mean (Table~\ref{tab:em-pixel}) reverses both: averaging over locations recovers the coverage and pays for it in false alarms.

\begin{table}[htbp]
  \centering
  \small
  \begin{tabular}{lccccc}
    \hline
    Model & RMSE $\downarrow$ & MAE $\downarrow$ & CSI $\uparrow$ & POD $\uparrow$ & FAR $\downarrow$ \\
    \hline
    Baseline               & \textbf{1.484}    & \textbf{0.408}    & \textbf{0.559}    & \textbf{0.884}    & 0.397 \\
    \hline
    RainFlow (AM)          & $1.980 \pm 0.019$ & $0.520 \pm 0.004$ & $0.455 \pm 0.003$ & $0.600 \pm 0.004$ & $\mathbf{0.347 \pm 0.002}$ \\
    RainFlow-Residual (AM) & $1.966 \pm 0.016$ & $0.527 \pm 0.004$ & $0.457 \pm 0.003$ & $0.609 \pm 0.004$ & $0.353 \pm 0.002$ \\
    \hline
  \end{tabular}
  \caption{\textbf{Pixel-wise skill of single members on the 2019 test set.} Baseline against average member (AM) of each generative model: each score is computed for every one of the 50 members and reported as mean $\pm$ standard deviation over members. RMSE and MAE in \si{\milli\metre\per\hour}; CSI, POD and FAR for the rain/no-rain event $RR \geq \SI{0.12}{\milli\metre\per\hour}$. Best per column in bold; arrows give the preferred direction.}
  \label{tab:pixel}
\end{table}

\clearpage

\appendix[C]

\appendixtitle{rain rate distribution: additional results}
\applabel{C}{app:marginal}

\begin{figure}[htbp]
  \centering
  \includegraphics[width=\linewidth]{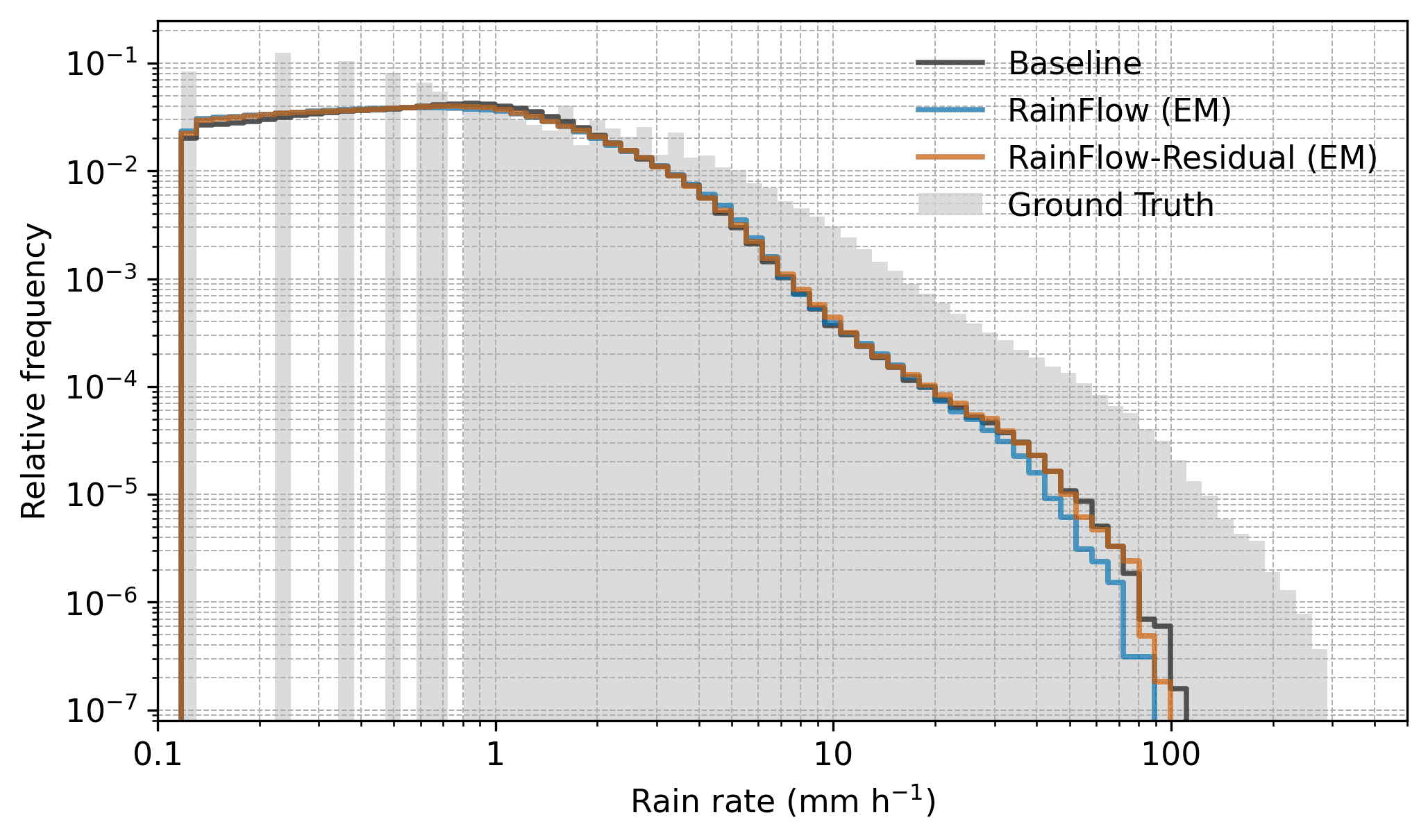}\\[2ex]
  \includegraphics[width=0.49\linewidth]{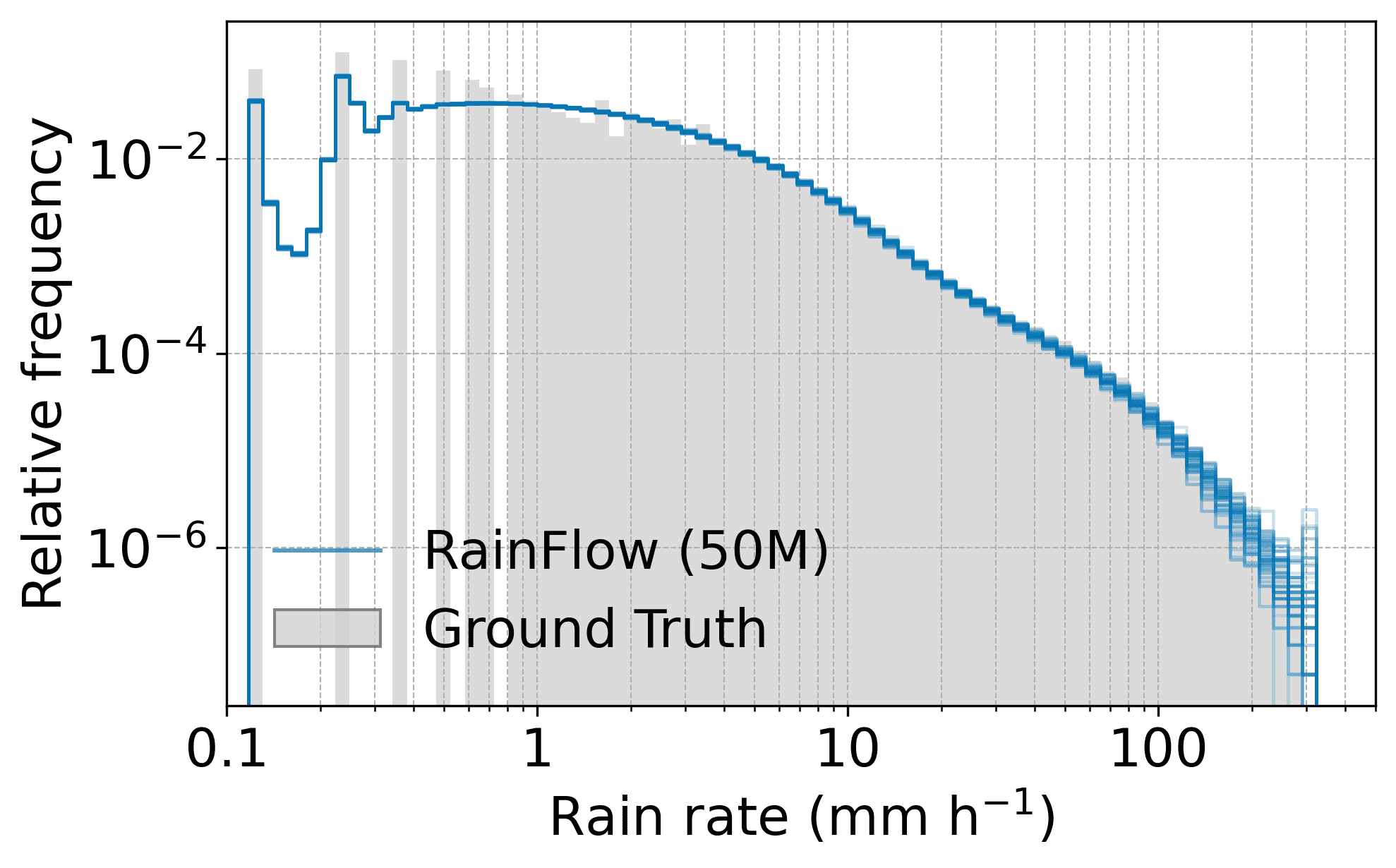}\hfill
  \includegraphics[width=0.49\linewidth]{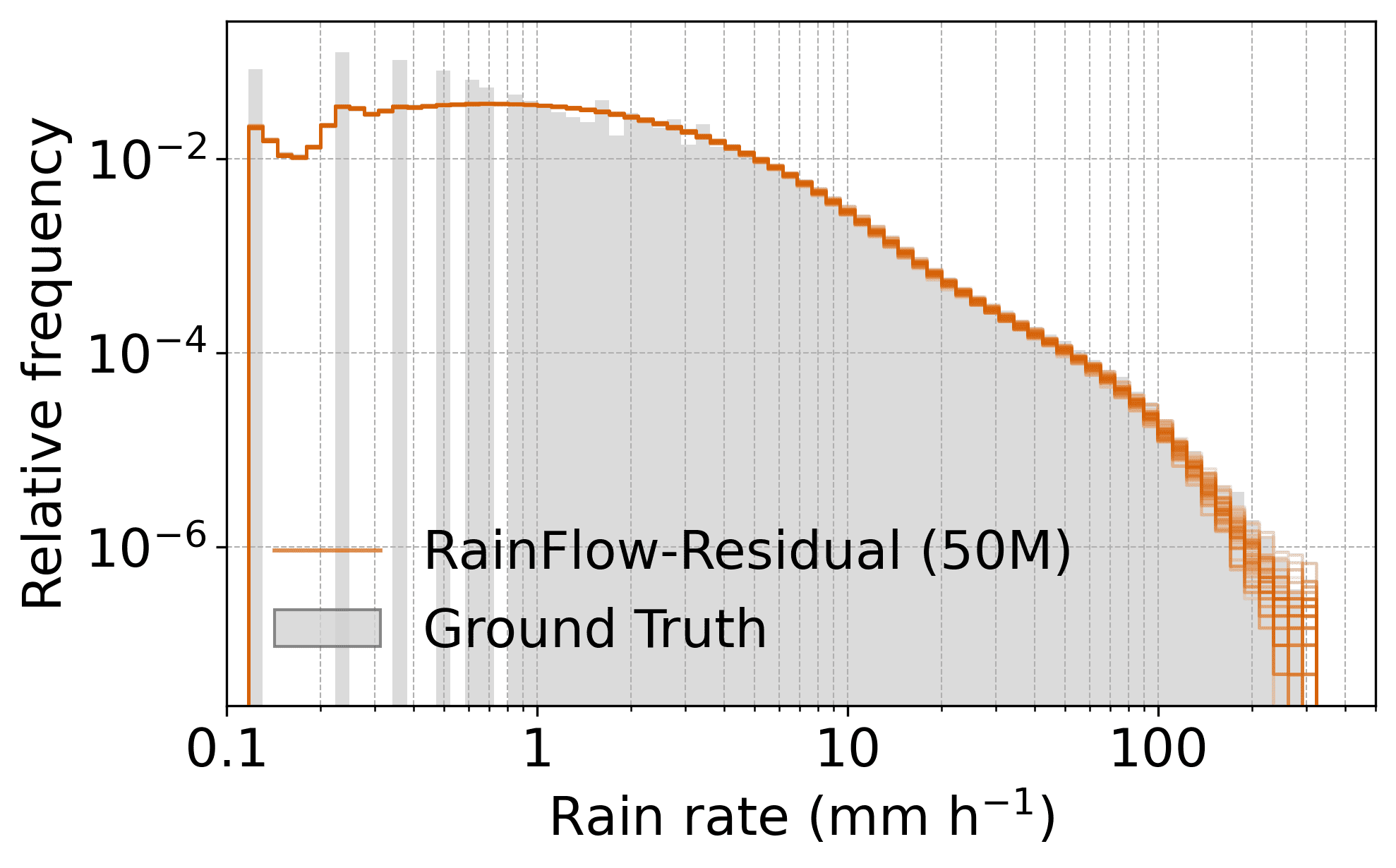}
\caption{\textbf{rain rate distribution of wet pixels ($RR \geq \SI{0.12}{\milli\metre\per\hour}$) for the ensemble mean and every member.} \emph{Top:} ensemble mean (EM) of each generative model against the baseline and the observations. Like the baseline, the ensemble means truncate near $\SI{100}{\milli\metre\per\hour}$: averaging removes the heavy tail that each member reproduces. \emph{Bottom:} all $50$ members of RainFlow (left) and RainFlow-Residual (right). The members are indistinguishable through the bulk of the distribution and differ only above about $\SI{100}{\milli\metre\per\hour}$, where pixel counts are small.}
  \label{fig:marginal-appendix}
\end{figure}

\clearpage
\appendix[D]

\appendixtitle{Power spectra: additional results}
\applabel{D}{app:psd}

\begin{figure}[htbp]
  \centering
  \includegraphics[width=\linewidth]{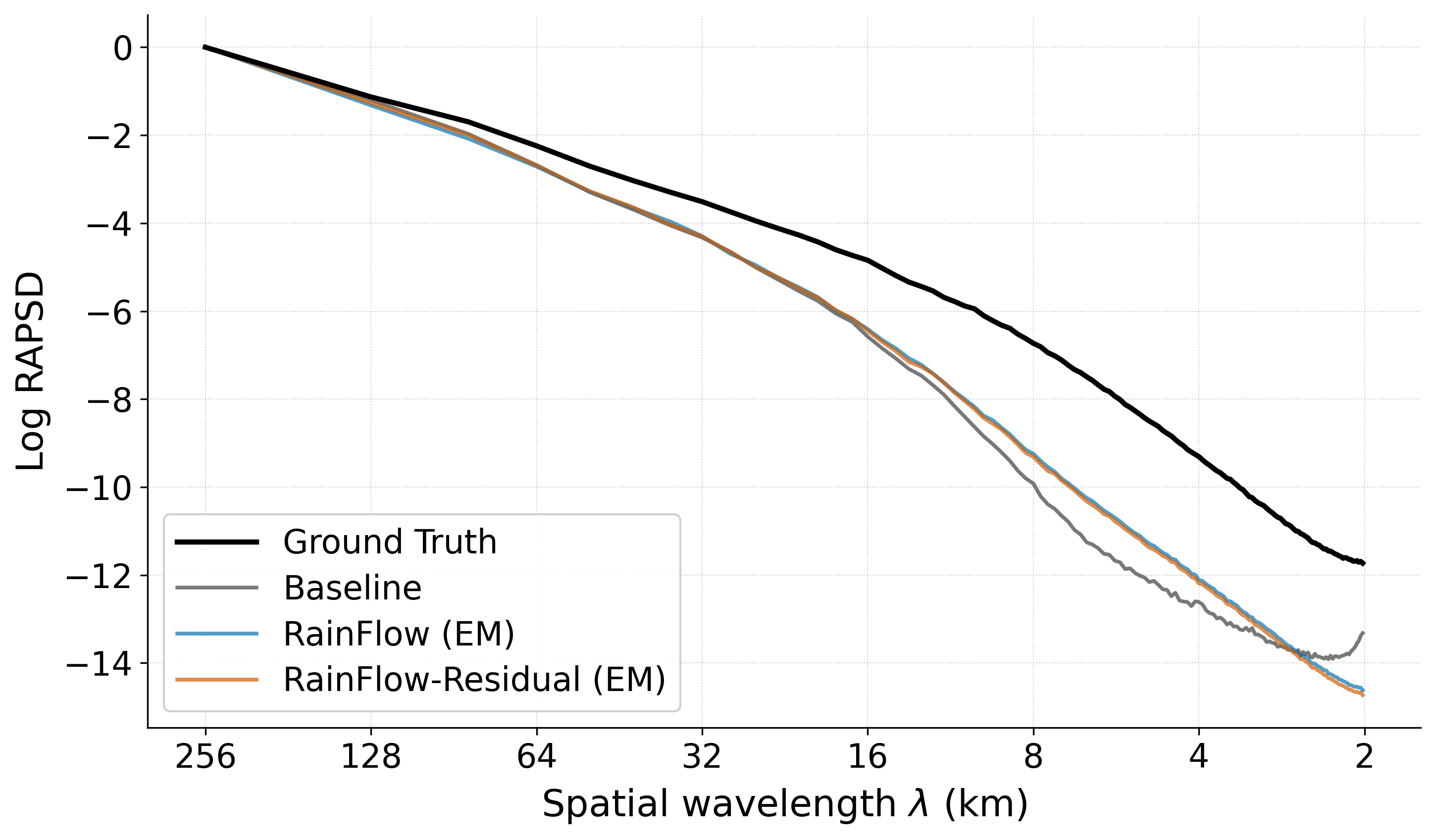}\\[2ex]
  \includegraphics[width=0.49\linewidth]{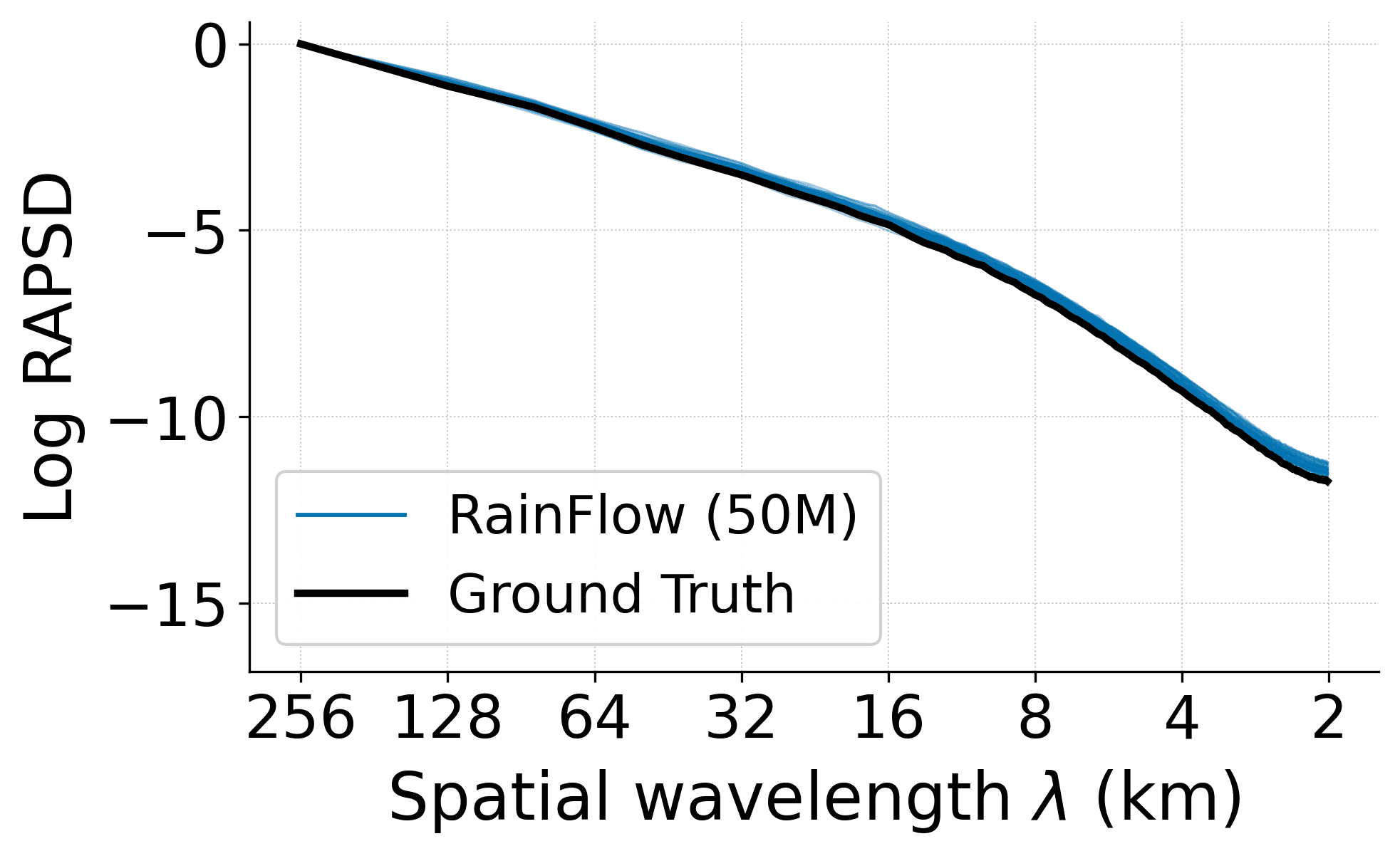}\hfill
  \includegraphics[width=0.49\linewidth]{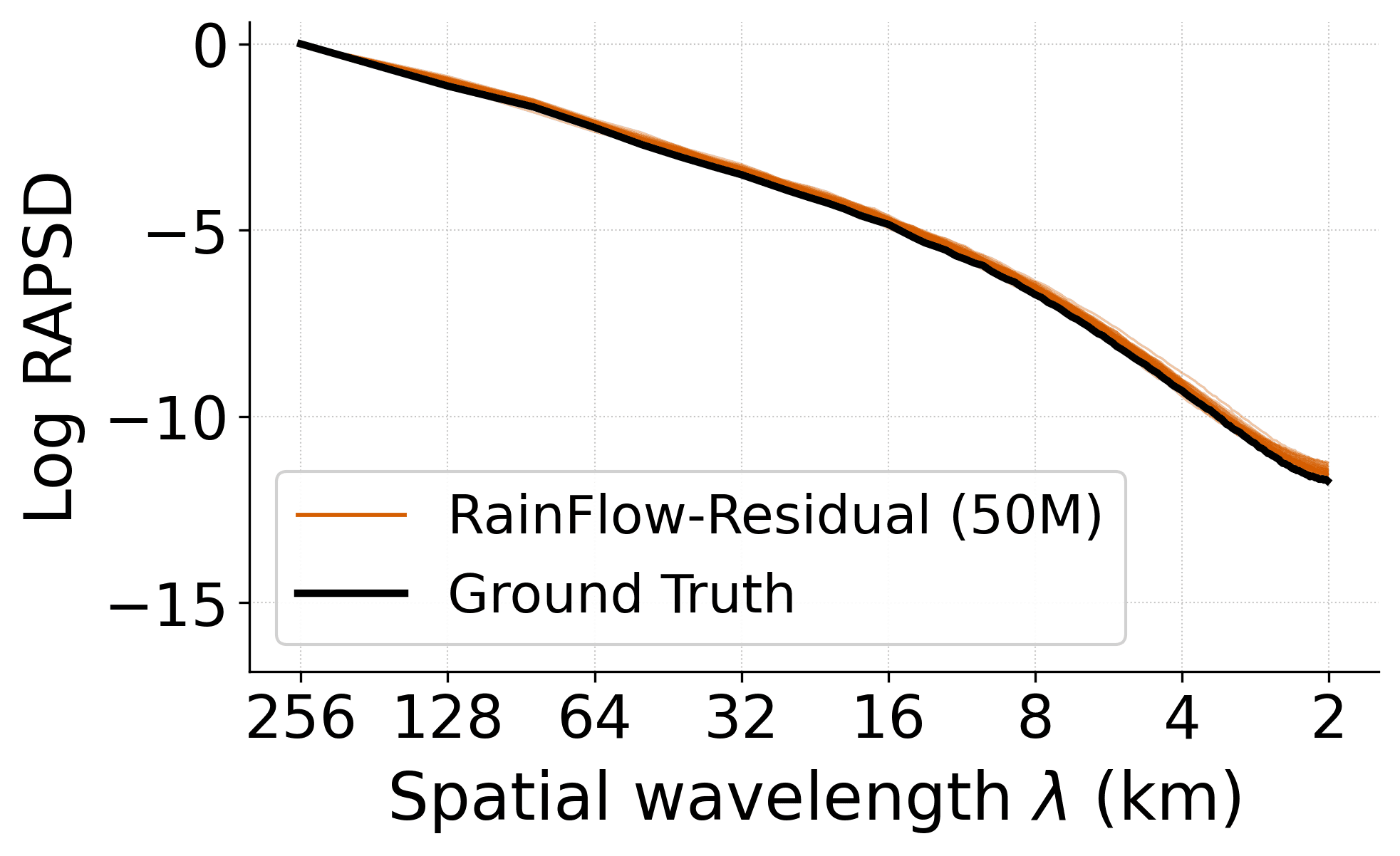}
  \caption{\textbf{Radially averaged power spectra of the ensemble mean and of every member.} \emph{Top:} ensemble mean (EM) of each generative model against the baseline and the observations. Like the baseline, the ensemble means lose power below roughly $\SI{100}{\kilo\metre}$. \emph{Bottom:} all $50$ members of RainFlow (left) and RainFlow-Residual (right). Their spectra are superimposed, so a single member (RM) is representative of the ensemble.}
  \label{fig:psd-appendix}
\end{figure}

\clearpage
\appendix[E]

\appendixtitle{Location skill: additional results}
\applabel{E}{app:location}

\begin{figure}[htbp]
  \centering
  \includegraphics[width=\linewidth]{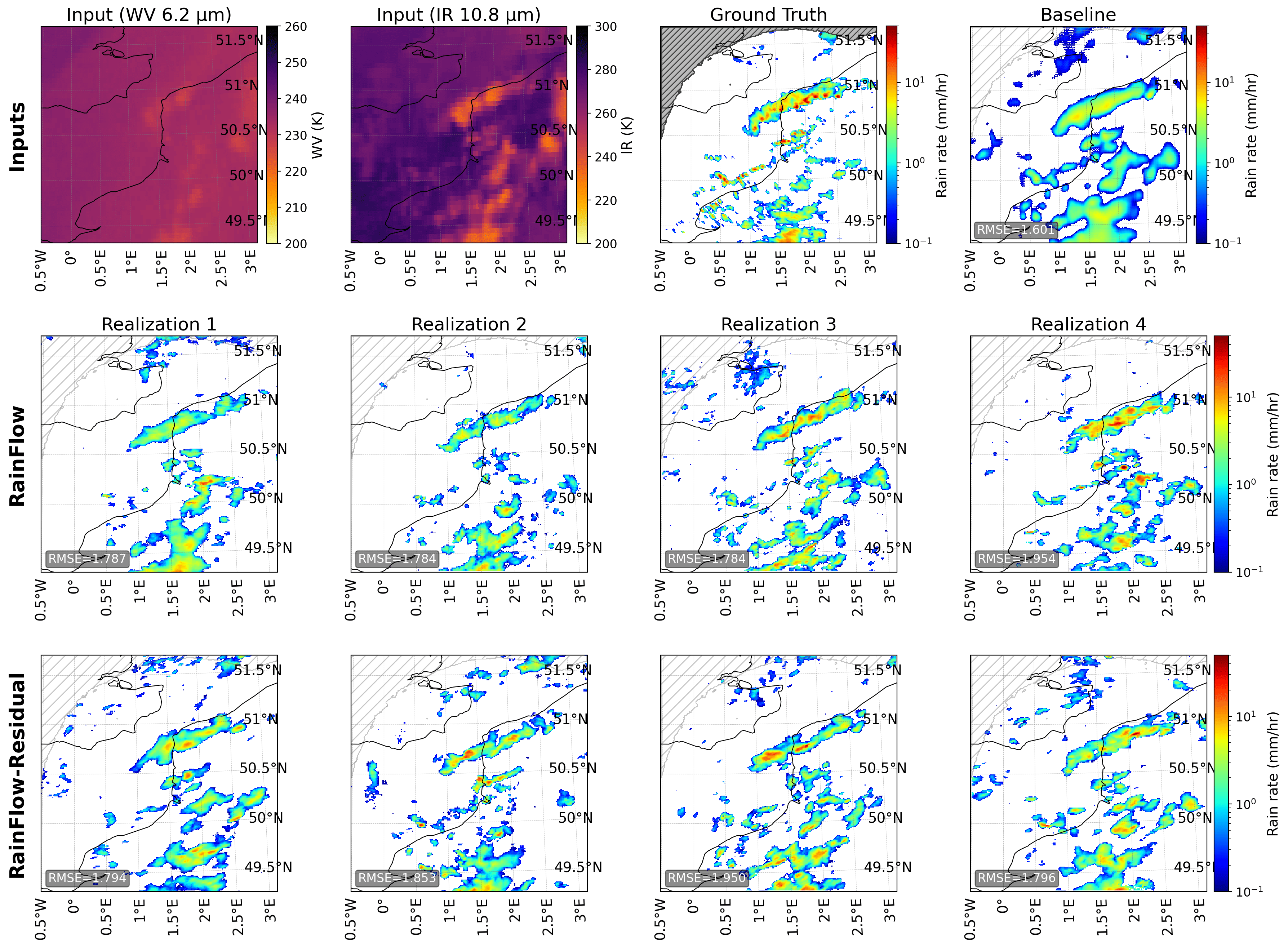}
  \caption{\textbf{Convective case study: four members of each ensemble.} Members of RainFlow and RainFlow-Residual for the scene of Fig.~\ref{fig:casestudy}, with the inputs, the radar observations and the baseline in the top row. Rain rate is on a logarithmic scale; the RMSE against the radar is annotated on each rain panel.}
  \label{fig:casestudies-appendix}
\end{figure}

\begin{figure}[htbp]
  \centering
  \includegraphics[width=\linewidth]{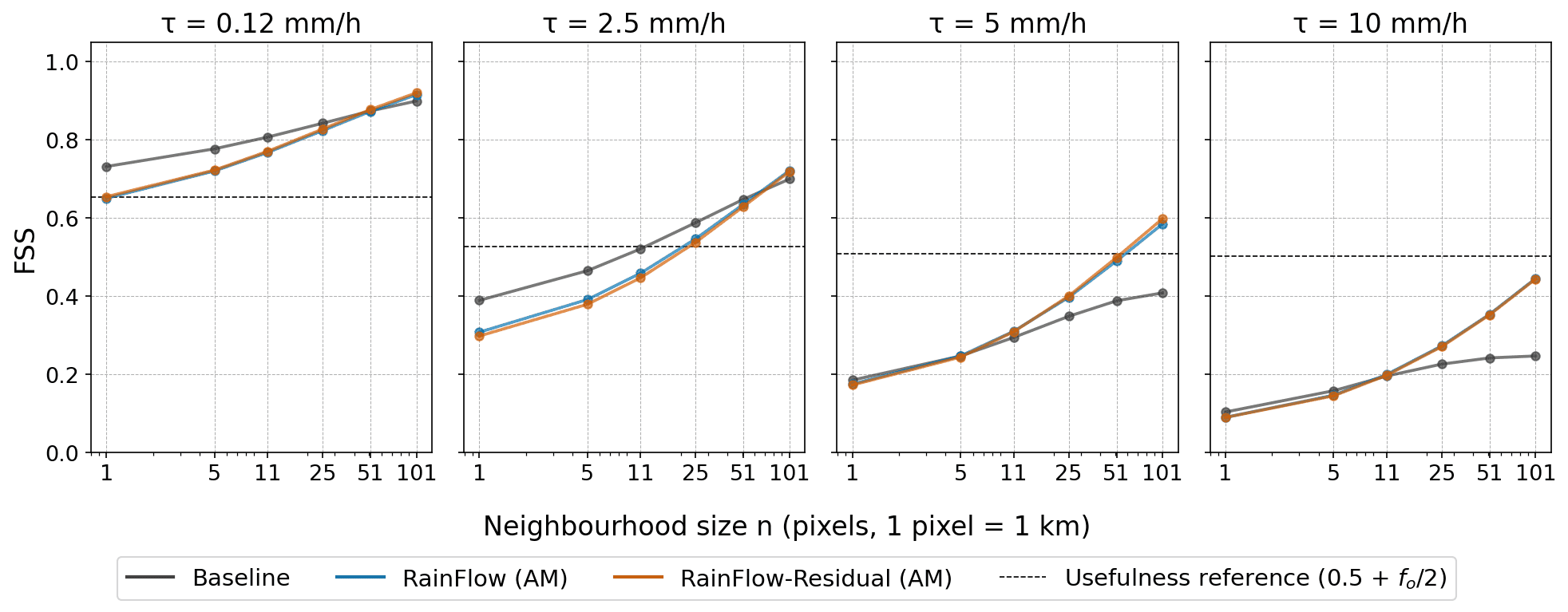}\\[2ex]
  \includegraphics[width=\linewidth]{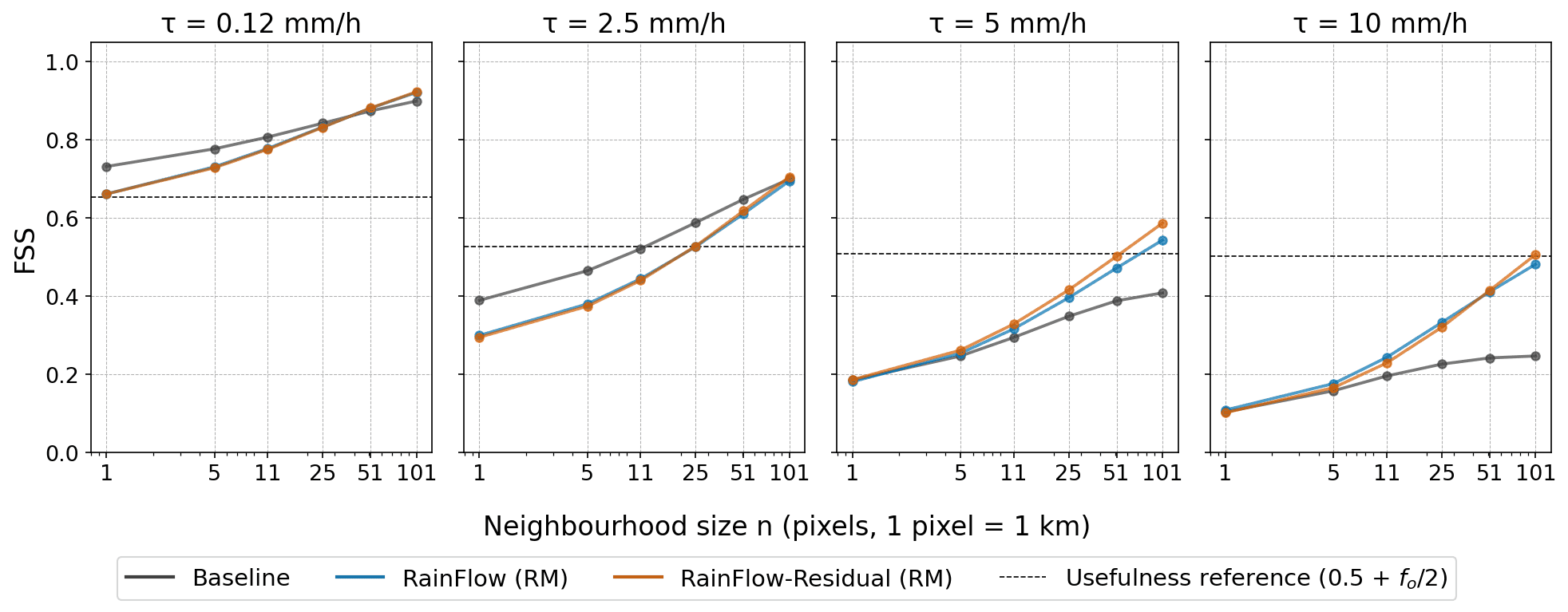}
  \caption{\textbf{Fractions skill score of single members.} \emph{Top:} classic FSS computed separately for each of the $50$ members and averaged over them (avFSS, AM). \emph{Bottom:} classic FSS of a single random member (RM). Both are shown against the FSS of the baseline, for the four thresholds of Fig.~\ref{fig:pfss}. Without pooling, a single member overtakes the baseline at large scales for light rain, from about $\SI{50}{\kilo\metre}$ at $0.12$ and $\SI{2.5}{\milli\metre\per\hour}$, and at progressively smaller scales as intensity increases, from about $\SI{10}{\kilo\metre}$ at $5$ and $\SI{10}{\milli\metre\per\hour}$. The single draw follows the average closely, with small departures at the highest thresholds.}
  \label{fig:fss-appendix}
\end{figure}

\clearpage

\appendix[F]

\appendixtitle{Conditioning on brightness temperature: observed wet pixels}
\applabel{F}{app:tb}

\begin{figure}[htbp]
  \centering
  \includegraphics[width=0.49\linewidth]{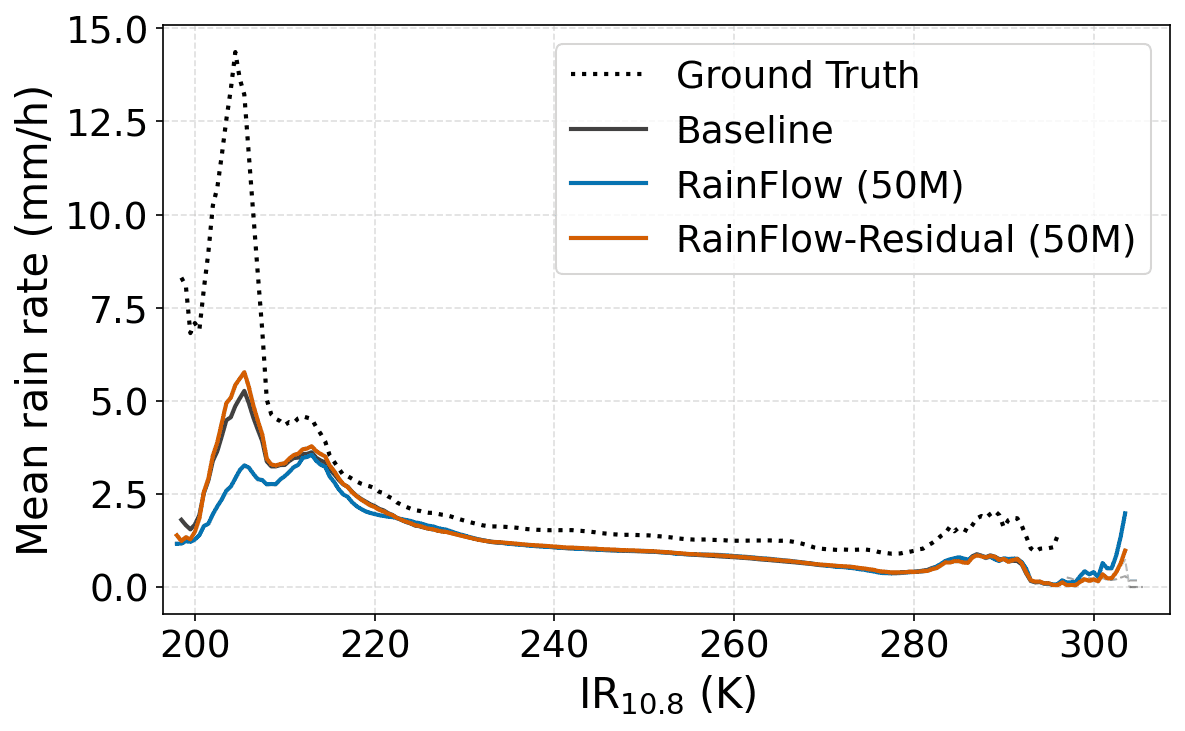}\hfill
  \includegraphics[width=0.49\linewidth]{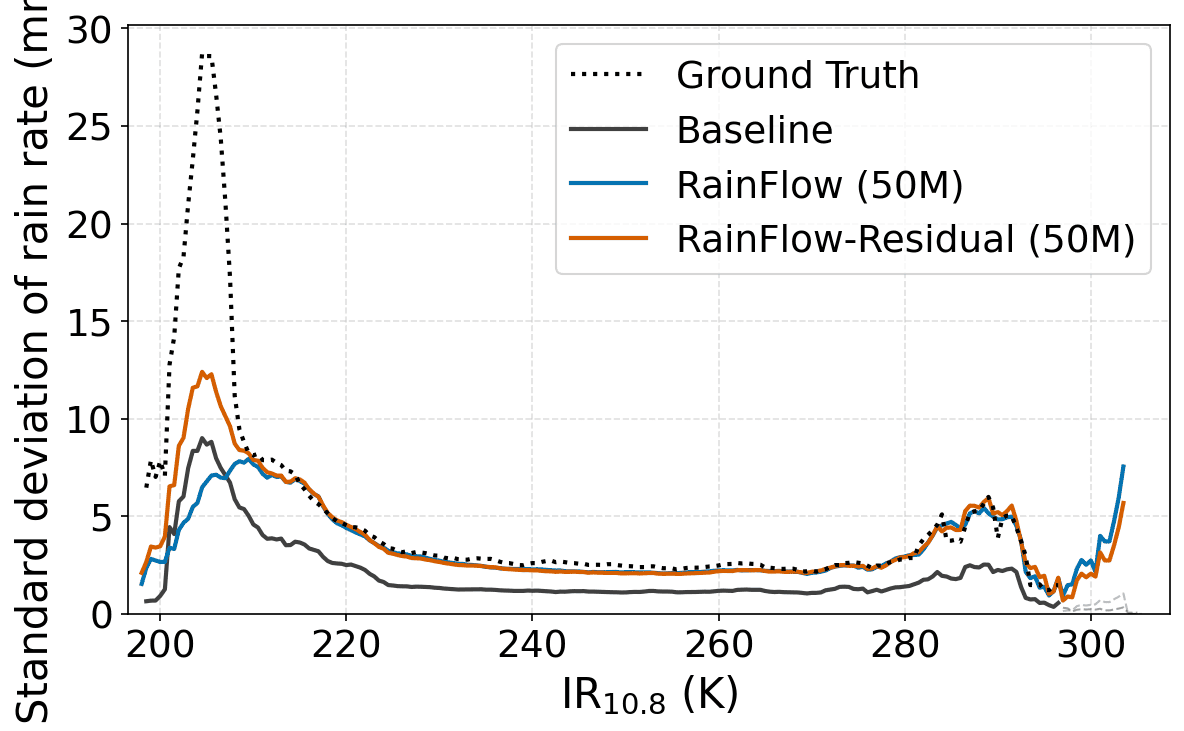}
  \caption{\textbf{Conditional moments of rain rate given $\mathrm{IR}_{10.8}$, observed wet pixels only.} As Fig.~\ref{fig:condtb}, restricted to pixels where the radar reports rain ($RR \geq \SI{0.12}{\milli\metre\per\hour}$, \S\ref{sec:data}). Mean (\emph{left}) and standard deviation (\emph{right}), against the observed values (dotted).}
  \label{fig:condtb-wet}
\end{figure}

Figure~\ref{fig:condtb-wet} repeats the conditional moments of \S\ref{sec:res-cond} on the pixels where the radar reports rain. This restriction selects the pixels using the observations, so a displaced field is scored at locations where it produced little rain, and the observed sample is wet by construction whereas the model sample at the same pixels is not. Both effects lower the model curves independently of any error in intensity.

The conditional mean is affected accordingly: all three models now lie below the observations at every brightness temperature, by roughly a third over most of the range, whereas on all pixels they agreed with the observations above about $\SI{215}{\kelvin}$. The conditional spread is not affected in the same way. Both ensembles follow the observed standard deviation from about $\SI{215}{\kelvin}$ to the warmest tops, including the secondary maximum near $\SI{288}{\kelvin}$, while the baseline remains at roughly half the observed value throughout. The conclusion of \S\ref{sec:res-cond} therefore holds under either selection: the ensembles reproduce the variability of the rain rate over most of the brightness temperature range, and underestimate it only under the coldest cloud tops.

\appendix[G]

\appendixtitle{Effect of the time-sampling density}
\applabel{G}{app:tsampling}

\begin{figure}[t]
  \centering
  \includegraphics[width=\linewidth]{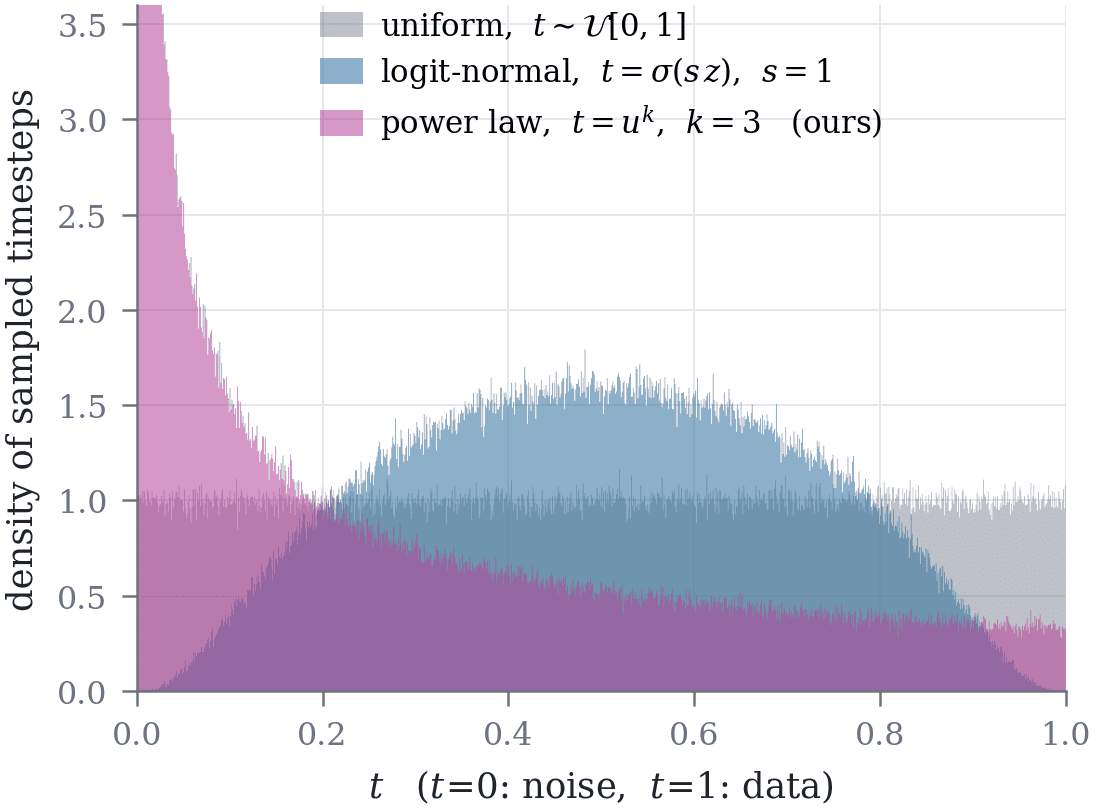}
  \caption{\textbf{Densities of the training time.} Uniform density of the original formulations, logit-normal density of \citet{esser2024scaling} and power law $t = u^{3}$ used in this work, with $t = 0$ the noise end of the trajectory and $t = 1$ the data end. They place $10\%$, $1.4\%$ and $46.4\%$ of training times below $t = 0.1$, respectively.}
  \label{fig:tdensity}
\end{figure}

This appendix compares the densities of training times discussed in \S\ref{sec:time-sampling} (Fig.~\ref{fig:tdensity}): the uniform density of the original formulations \citep{lipman2023flow, tong2024improving}; the logit-normal density of \citet{esser2024scaling}, $t = 1/(1 + e^{-z})$ with $z \sim \mathcal{N}(0, 1)$; and the power law $t = u^{k}$ with $u \sim \mathcal{U}[0,1]$ and $k = 3$, used in the main text. All other settings are those of \S\ref{sec:training-sampling}. Table~\ref{tab:tsampling} reports the scores of each configuration on the test set, with $95\%$ bootstrap intervals over test scenes.

\begin{table}[t]
  \centering
  \small
  \setlength{\tabcolsep}{3.5pt}
  \begin{tabular}{lcccccc}
    \hline
    $p(t)$ & $\Pr(t{<}0.1)$ & CRPS $\downarrow$ & twCRPS$_{\tau = 10}$ $\downarrow$ & RMSE (EM) $\downarrow$ & SSR $\to 1$ & pFSS $\uparrow$ \\
    \hline
    \multicolumn{7}{l}{\textit{RainFlow}} \\
    Uniform                    & $10\%$   & $0.285$ & $3.057$ & $1.533$ & $0.878$ & $0.755$ \\
                               &          & {\scriptsize$[0.270,\,0.302]$} & {\scriptsize$[2.542,\,3.686]$} & {\scriptsize$[1.404,\,1.683]$} & {\scriptsize$[0.806,\,0.951]$} & {\scriptsize$[0.743,\,0.767]$} \\
    Logit-normal               & $1.4\%$  & $0.308$ & $3.162$ & $1.614$ & $\mathbf{1.092}$ & $0.699$ \\
                               &          & {\scriptsize$[0.289,\,0.327]$} & {\scriptsize$[2.631,\,3.803]$} & {\scriptsize$[1.480,\,1.766]$} & {\scriptsize$[1.004,\,1.184]$} & {\scriptsize$[0.688,\,0.708]$} \\
    Power law, $k = 3^\dagger$ & $46.4\%$ & $\mathbf{0.275}$ & $\mathbf{3.009}$ & $\mathbf{1.495}$ & $0.887$ & $\mathbf{0.779}$ \\
                               &          & {\scriptsize$[0.260,\,0.290]$} & {\scriptsize$[2.504,\,3.633]$} & {\scriptsize$[1.368,\,1.641]$} & {\scriptsize$[0.815,\,0.958]$} & {\scriptsize$[0.767,\,0.791]$} \\
    \hline
    \multicolumn{7}{l}{\textit{RainFlow-Residual}} \\
    Uniform                    & $10\%$   & $0.279$ & $\mathbf{3.000}$ & $1.491$ & $\mathbf{0.938}$ & $0.779$ \\
                               &          & {\scriptsize$[0.265,\,0.294]$} & {\scriptsize$[2.517,\,3.582]$} & {\scriptsize$[1.380,\,1.614]$} & {\scriptsize$[0.883,\,0.995]$} & {\scriptsize$[0.767,\,0.790]$} \\
    Logit-normal               & $1.4\%$  & $0.282$ & $3.026$ & $1.515$ & $1.251$ & $0.768$ \\
                               &          & {\scriptsize$[0.267,\,0.297]$} & {\scriptsize$[2.549,\,3.604]$} & {\scriptsize$[1.407,\,1.636]$} & {\scriptsize$[1.182,\,1.321]$} & {\scriptsize$[0.757,\,0.780]$} \\
    Power law, $k = 3^\dagger$ & $46.4\%$ & $\mathbf{0.278}$ & $3.001$ & $\mathbf{1.487}$ & $0.882$ & $\mathbf{0.781}$ \\
                               &          & {\scriptsize$[0.263,\,0.293]$} & {\scriptsize$[2.517,\,3.583]$} & {\scriptsize$[1.376,\,1.611]$} & {\scriptsize$[0.829,\,0.936]$} & {\scriptsize$[0.769,\,0.792]$} \\
    \hline
  \end{tabular}
  \caption{\textbf{Effect of the time-sampling density on the 2019 test set.} Densities are characterised by the fraction of training times drawn below $t = 0.1$, at the noise end of the trajectory. CRPS and twCRPS above $\SI{10}{\milli\metre\per\hour}$ of the full ensemble (E); RMSE of the ensemble mean (EM); spread--skill ratio (SSR); pFSS for the rain/no-rain event ($\tau = \SI{0.12}{\milli\metre\per\hour}$) at the pixel scale, all defined in App.~\ref{app:evaluation-metrics}. CRPS and RMSE are in \si{\milli\metre\per\hour}, twCRPS in $10^{-2}$\,\si{\milli\metre\per\hour}. Arrows indicate whether lower ($\downarrow$) or higher ($\uparrow$) is better; for the SSR, the closer to 1 the better. The best value for each model is in bold. Brackets give $95\%$ bootstrap intervals over the 1349 test scenes. The configurations marked $\dagger$ are those used in \S\ref{sec:results}.}
  \label{tab:tsampling}
\end{table}

For RainFlow, the scores order with the weight that the density places near the noise end. The logit-normal density, which draws only $1.4\%$ of training times below $t = 0.1$, gives a CRPS $12\%$ above that of the power law, and the uniform density $3.6\%$ above it. The same ordering holds for the RMSE of the ensemble mean and for location skill: the pFSS for the rain/no-rain event at the pixel scale is $0.70$ under the logit-normal density, $0.75$ under the uniform density and $0.78$ under the power law. The logit-normal density also produces an overdispersed ensemble ($\mathrm{SSR} = 1.09$, with an interval above 1), with a larger spread and a less accurate mean. This is consistent with the endpoint argument of \S\ref{sec:time-sampling}: when training rarely visits the noise end, the network learns less precisely how the initial velocity depends on the conditioning, and the samples are less tied to it, hence wider and less well centred. These differences exceed the gap between the two models: with the power law, RainFlow scores $1.2\%$ better than RainFlow-Residual, whereas it falls $2.2\%$ behind under the uniform density and $9.2\%$ behind under the logit-normal density.

RainFlow-Residual is much less sensitive. Its CRPS varies by $1.4\%$ across the three densities, and by $0.4\%$ between the uniform density and the power law, against $3.6\%$ for RainFlow, as expected when the baseline prediction supplies the conditional mean that RainFlow must learn at the noise end (\S\ref{sec:time-sampling}). The density nonetheless affects its dispersion: the spread--skill ratio is $0.88$ with the power law and $0.94$ with the uniform density, but $1.25$ with the logit-normal density, whose interval lies well above the other two. The residual formulation thus makes the skill of the ensemble robust to the time density, but not its spread.

The intervals in Table~\ref{tab:tsampling} account for the sampling of test scenes only; each configuration is a single training run, so run-to-run variability is not included. The intervals of different configurations also overlap for most scores, because they contain the scene-to-scene variability shared by all of them: scenes with extensive rain yield larger errors for every configuration. They bound the score of each configuration rather than the difference between two of them, and only the largest contrasts, such as the pFSS of RainFlow and the spread--skill ratios under the logit-normal density, are resolved by them alone. Differences of about $1\%$, such as those among the configurations of RainFlow-Residual, should therefore not be over-read; the larger differences found for RainFlow are consistent across the probabilistic, pixel-wise and location scores.

\clearpage

\bibliographystyle{ametsocV6}   
\bibliography{bib}

\end{document}